\documentclass[article]{jss}

\usepackage{thumbpdf,lmodern}

\usepackage{framed}
\usepackage{pifont}
\usepackage{multirow}
\usepackage{booktabs}
\usepackage{rotating}
\usepackage{placeins}
\usepackage{amsmath,amssymb}

\author{Robin Denz\\Ruhr-University of Bochum
   \And Nina Timmesfeld\\Ruhr-University of Bochum}
\Plainauthor{Robin Denz, Nina Timmesfeld}

\title{Simulating Complex Crossectional and Longitudinal Data using the \pkg{simDAG} \proglang{R} Package}
\Plaintitle{Simulating Complex Crossectional and Longitudinal Data using the simDAG R Package}
\Shorttitle{The \pkg{simDAG} \proglang{R} Package}

\Abstract{
Generating artificial data is a crucial step when performing Monte-Carlo simulation studies. Depending on the planned study, complex data generation processes (DGP) containing multiple, possibly time-varying, variables with various forms of dependencies and data types may be required. Simulating data from such DGP may therefore become a difficult and time-consuming endeavor. The \pkg{simDAG} \proglang{R} package offers a standardized approach to generate data from simple and complex DGP based on the definition of structural equations in directed acyclic graphs using arbitrary functions or regression models. The package offers a clear syntax with an enhanced formula interface and directly supports generating binary, categorical, count and time-to-event data with arbitrary dependencies, possibly non-linear relationships and interactions. It additionally includes a framework to conduct discrete-time based simulations which allows the generation of longitudinal data on a semi-continuous time-scale. This approach may be used to generate time-to-event data with both recurrent or competing events and possibly multiple time-varying covariates, which may themselves have arbitrary data types. In this article we demonstrate the vast amount of features included in \pkg{simDAG} by replicating the DGP of multiple real Monte-Carlo simulation studies.
}

\Plainkeywords{simulation, time-to-event, structural equation, directed acyclic graph, longitudinal data, discrete-time, R}

\Address{
  Robin Denz, Nina Timmesfeld\\
  Department of Medical Informatics, Biometry and Epidemiology\\
  Ruhr-University of Bochum\\
  Universit\"atsstr.~105\\
  D-44789 Bochum, Germany\\
  E-mail: \email{robin.denz@rub.de}\\
  URL: \url{http://www.amib.ruhr-uni-bochum.de/index.php}
}

\usepackage{Sweave}
\begin{document}

\sloppy

\section{Introduction} \label{chap::introduction}

\subsection{Motivation}

Applied researchers and statisticians frequently use Monte-Carlo simulation techniques in a variety of ways. They are used to estimate required sample sizes \citep{Arnold2011}, formally compare different statistical methods \citep{Morris2019, Denz2023}, help design and plan clinical trials \citep{Kimko2002, Nance2024} or for teaching purposes \citep{Sigal2016, Fox2022}. The main reason for their broad usage is that the researcher has full control over the true data generation process (DGP). In general, the researcher will define a DGP appropriate to the situation and generate multiple datasets from it. Some statistical analysis technique is then applied to each dataset and the results are analyzed. A crucial step in every kind of Monte-Carlo simulation study is thus the generation of these datasets.
\par
Depending on the DGP that is required by the researcher, this step may become very difficult and time consuming. For example, some Monte-Carlo simulations require the generation of complex longitudinal data with variables of different types that are causally related in various ways \citep{Asparouhov2020}. Some of these possible data types are continuous variables, categorical variables, count variables or time-to-event variables. All of these require different parametrizations and simulation strategies. If interactions or non-linear relationships between these variables are required, simulating data from the DGP becomes even more challenging.
\par
Generating any artificial data requires (1) a formal description of the DGP, (2) the knowledge of an algorithm that may be used to generate the data from this DGP, and (3) the ability to create a software application to implement that algorithm. Although many statisticians may have no problem with theses steps, this might not be the case for more applied researchers. More importantly, the third step in particular may require a high level of expertise in programming, because the resulting program has to be validated extensively while it also has to be computationally efficient enough to allow potentially thousands of datasets to be generated in a reasonable amount of time. Additionally, it also has to be flexible enough to allow the user to easily make changes to the DGP to be useful in most cases \citep{Sofrygin2017}. A comprehensive software application that automates most of the required work would therefore be of great benefit to the scientific community.
\par
In this article we present the \pkg{simDAG} \proglang{R} package, which offers an easy to use and consistent framework to generate arbitrarily complex crossectional and longitudinal data. The aim of the package is to make all three steps of the data generation process easier by giving users a standardized way to define the desired DGP, which can then be used directly to generate the data without further user input. It does so by requiring the user to define a directed acyclic graph (DAG) with additional information about the associations between the supplied variables \citep{Pearl2009}. The package was created using the \proglang{R} programming language \citep{RCT2024} and is available on the Comprehensive \proglang{R} Archive Network (CRAN) at \url{https://cran.r-project.org/package=simDAG}.

\subsection{Using DAGs to define data generation processes}

\begin{figure}[!htb]
	\centering
	\includegraphics[width=0.4\linewidth]{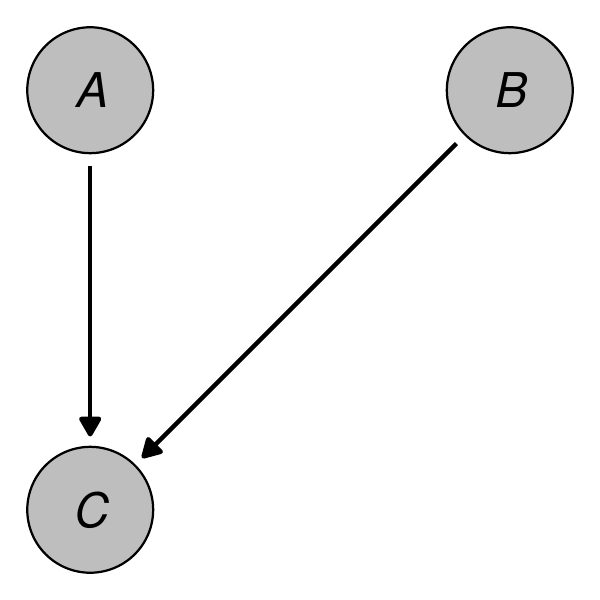}
	\caption{An example DAG with three nodes.}
	\label{fig::example_dag}
\end{figure}

In this package, the user is required to describe the desired DGP as a causal DAG. Formally, a DAG is a mathematical graph consisting of a set of $V$ nodes (or vertices) and a set of $E$ edges (or links) connecting pairs of nodes. As its' name suggests, a DAG consists only of \emph{directed} edges and is \emph{acyclic}, meaning that there are no cycles when following directed paths on the DAG \citep{Byeon2023}. A causal DAG is a special sort of DAG in which the nodes represent random variables and the edges represent directed causal relationships between these variables \citep{Pearl2009}. A very simple example containing only three nodes and no time-dependencies is given in Figure~\ref{fig::example_dag}. The DAG in this figure contains a directed arrow from $A$ to $C$ and from $B$ to $C$. This translates to the assumptions that there is a direct causal effect of $A$ on $C$ and of $B$ on $C$, but no direct causal relationship between $A$ and $B$ (due to the absence of an arrow between them).

Such DAGs are the cornerstone of the \emph{structural approach} to causal inference developed by \citet{Pearl2009} and \citet{Spirtes2000}. They are used extensively in social research \citep{Wouk2019}, econonomics \citep{Imbens2020} and epidemiology \citep{Byeon2023} to encode causal assumptions about the real underlying DGP of empirical data. For empirical research such graphs are very useful because they give a clear overview of the causal assumptions made by the researchers. By using causal graphical methods such as the \emph{backdoor} criterion \citep{Pearl2009} or the \emph{frontdoor} criterion \citep{Pearl1995}, it is also possible to use such graphs to determine which variables need to be adjusted for in order to get unbiased estimates of certain causal effects. The \pkg{daggitty} \proglang{R} package directly implements multiple tools for this kind of usage \citep{Textor2016}.
\par
These kind of DAGs can be formally described using \emph{structural equations}. These equations describe how each node is distributed. For example, a general set of structural equations that may be used to describe the DAG in Figure~\ref{fig::example_dag} are:

\begin{equation}
	\begin{aligned}
		A \sim & f_A(U_A), \\
		B \sim & f_B(U_B), \\
		C \sim & f_C(A, B, U_C). \\
	\end{aligned}
\end{equation}

\par
In these equations, the unspecified functions $f_A$, $f_B$ and $f_C$ describe how exactly the nodes are distributed, possibly conditional on other nodes. The terms $U_A$, $U_B$ and $U_C$ denote random errors or disturbances. If the functions in these structural equations are not specified and some assumption on the probability distribution of the error terms is made, this is equivalent to a non-parametric structural equation model \citep{Pearl2009, Sofrygin2017}.
\par
To make the generation of data from a DAG possible, however, it is not enough to only specify which variables are causally related to one another. The structural equations now also have to be fully specified. This means that the distribution functions of each node will have to be defined in some way by the user. A popular way to do this is to use regression models and parametric distributions \citep{Kline2023}, but in theory any kind of function may be used, allowing the definition of arbitrarily complex DGPs. Continuing the example from above, we could define the structural equations of the DAG as follows:

\begin{equation} \label{eq::structural_eq_example}
	\begin{aligned}
		A \sim & N(0, 1), \\
		B \sim & N(0, 1), \\
		C \sim & -2 + A\cdot0.3 + B\cdot-2 + N(0, 1). \\
	\end{aligned}
\end{equation}

This means that both $A$ and $B$ are independent standard normally distributed variables and that $C$ follows a simple linear regression model based on $A$ and $B$ with an independent normally distributed error term with mean zero. Once all structural equations and distribution functions have been defined, data may be generated from the DAG using a fairly simple algorithm. This algorithm essentially generates data for one node at a time, using only the supplied definitions and the data generated in previous steps. This step-wise method relies on the fact that every DAG can be \emph{topologically sorted}, which means that there is always an ordering of the nodes such that for every link $(u_i, u_j)$ between nodes $u_i$ and $u_j$, $u_i$ comes before $u_j$ \citep{Kahn1962}.
\par
The generation of the data starts by ordering the nodes of the graph in such a topologically sorted way. This means that nodes in the DAG that have no arrows pointing into them, sometimes called \emph{root nodes}, are processed first. Data for these kinds of nodes can be generated by sampling from a pre-specified parametric distribution, such as a Gaussian distribution or a beta distribution, or through any other process, such as re-sampling based strategies \citep{Carsey2014}. Once data for these nodes has been generated, the data for the next node in the topological order will be generated, based on the already generated data. These nodes are called \emph{child nodes}, because they are dependent on other nodes, which are called their \emph{parent nodes} \citep{Byeon2023}. For the example DAG shown earlier, the two possible topological sortings are:

\begin{equation}
	(A, B, C) \quad \text{and} \quad (B, A, C).
\end{equation}

Here, both $A$ and $B$ are root nodes because they do not have any parents and $C$ is a child node of both of its' parents $A$ and $B$. To generate data for this example using the algorithm described above, one would first generate $n$ random draws from a standard normal distribution for both $A$ and $B$. Next, one would calculate the linear combination of these values as specified by the linear regression model in Equation~\ref{eq::structural_eq_example} and add $n$ random draws from another standard normal distribution to it (which represents the error term). In \proglang{R}, this simple example could be simulated using the following code:

\begin{Schunk}
\begin{Sinput}
R> set.seed(43)
R> n <- 100
R> A <- stats::rnorm(n)
R> B <- stats::rnorm(n)
R> C <- -2 + A*0.3 + B*-2 + stats::rnorm(n)
\end{Sinput}
\end{Schunk}

Although the manual code required for this example is fairly simple, this is no longer the case in DAGs with more nodes and/or a more complex DGP (for example one including different data types). The \pkg{simDAG} package offers a standardized way to define any possible DAG and the required distribution functions to facilitate a clear and reproducible workflow.
\par
The previous explanations and the given example focused on the simple case of crossectional data without any time-dependencies. It is, however, fairly straightforward to include a time-varying structure into any DAG as well by simply adding a time-index to the time-varying nodes and repeating the node for each point in time that should be considered \citep{Hernan2020}. The proposed package features computationally efficient functions to automate this process for large amounts of time-points using a \emph{discrete-time simulation} approach \citep{Tang2020}. Although this procedure relies on a discrete time scale, it can be used to generate data on a semi-continuous time-scale by using very small steps in time. This is described in more detail in Section~\ref{chap::discrete_time_sim}.
\par
Note also that while causal DAGs imply a specific causal structure, the algorithms and code described here do not necessitate that this causal structure has to be interpreted as such in the generated data. For example, the structural equations in Equation~\ref{eq::structural_eq_example} state that $A$ and $B$ are direct causes of $C$, but the datasets that can be generated from these equations could also be interpreted as $A$ and $B$ being only associated with $C$ for unknown reasons. As long as the desired DGP can be \emph{described} as a DAG, which is almost always the case, this strategy may be used effectively to generate data even for Monte-Carlo studies \emph{not} concerned with causal inference.
\par
Although the data-generation algorithm described above is appropriate for most applications, it may not be the best choice for validating causal discovery methods, due to the marginal variance of each variable increasing along the order of the topological sorting \citep{Reisach2021}. Other methods, such as the onion method proposed by \citet{Andrews2024} may be preferable in this particular case.

\subsection{Comparison with existing software}

There are many different software packages that may be used to generate data in the \proglang{R} programming language and other languages such as \proglang{Python}. It is infeasible to summarise all of them here, so we will only focus on a few that offer functionality similar to the \pkg{simDAG} package instead. The following review is therefore not intended to be exhaustive. We merely aim to show how the existing software differs from the proposed package.
\par
Multiple \proglang{R} packages support the generation of synthetic data from fully specified structural equation models. The \pkg{lavaan} package \citep{Rosseel2012}, the \pkg{semTools} package \citep{Jorgensen2022} and the \pkg{simsem} package \citep{Pornprasertmanit2021} are a few examples. However, these packages focus soley on structural equation models with linear relationships and as such do not allow the generation of data with different data types. For example, none of these packages allow the generation of time-to-event data, which is a type of data that is often needed in simulation studies. Specialized \proglang{R} packages such as the \pkg{survsim} \citep{Morina2017}, \pkg{simsurv} \citep{Brilleman2021}, \pkg{rsurv} \citep{Demarqui2024} and \pkg{reda} \citep{Wang2022} packages may be used to simulate such data instead. Although some of these packages allow generation of recurrent events, competing events and general multi-state time-to-event data, unlike the \pkg{simDAG} package, none of them support arbitrary mixtures of these data types or time-varying covariates.
\par
Other packages, such as the \pkg{simPop} package \citep{Templ2017} and the \pkg{simFrame} package \citep{Alfons2010} allow generation of more complex synthetic data structures as well, but are mostly focused on generating data that mimicks real datasets. Similarly, the \pkg{simtrial} package offers very flexible tools for the generation of randomized controlled trial data, but it would be difficult to use it to generate other data types. Software directly based on causal DAGs as DGPs also exists. Although it is not stated in the package documentation directly, the \pkg{simstudy} package \citep{Goldfeld2020} also relies on the DAG based algorithm described earlier. It supports the use of different data types and custom generation functions, but only has partial support for generation of longitudinal data. Alternatively, the \proglang{Python} library \pkg{DagSim} \citep{Hajj2023} allows users to generate arbitrary forms of data, while also allowing the user to supply custom functions for the data generation process. The price for this flexibility is, however, that not many default options are implemented in the library.
\par
Finally, the \pkg{simcausal} \proglang{R} package \citep{Sofrygin2017} is very similar to the \pkg{simDAG} package and was in fact a big inspiration for it. Like the \pkg{simDAG} package, it is also based on the causal DAG framework. The syntax for defining a DAG is nearly the same, with some differences in how formula objects can and should be specified. Unlike the proposed package, however, the \pkg{simcausal} package is focused mostly on generating data for simulation studies dealing with causal inference. As such, it also directly supports the generation of data after performing some interventions on the DAG \citep{Pearl2009}. Although the proposed package lacks such functionality, it is a lot more flexible in terms of what data can be generated. \pkg{simDAG} supports the use of arbitrary data generation functions, the definition of interactions, non-linear relationships and mixed model syntax in its' formula interface and categorical input data for nodes. None of these features are present in \pkg{simcausal} \citep{Sofrygin2017}.

\subsection{Organization of this article}

First, we will introduce the most important functionality of the proposed package by describing the core functions and the usual workflow when employing the package. This includes a detailed description of how to translate the theoretical description of the desired DGP into a \code{DAG} object, which may then be used to generate data. Afterwards we will illustrate the capabilities of the package by reproducing the DGPs of multiple real Monte-Carlo simulation studies. Two DGPs describing the generation of crossectional data and longitudinal data with only few considered points in time will be considered first. Afterwards, an introduction into the generation of more complex longitudinal data utilizing the discrete-time simulation approach is presented. Finally, the package and its potential usefulness is discussed.

\newpage

\section{The workflow}

\subsection{Included functions}

The following functions are used in a typical workflow using the \pkg{simDAG} \proglang{R} package.

\begin{description}
	\item[{\code{empty\_dag()}}] initializes an empty \code{DAG} object, which should be later filled with information on relevant nodes. \code{DAG} objects are the most important data structure of this package. How to define and use them is illustrated in more detail below.

	\item[{\code{node()} and \code{node\_td()}}] can be used to define one or multiple nodes each. These functions are typically used to fill the \code{DAG} objects with information about how the respective node should be generated, e.g., which other nodes it depends on (if any), whether it is time-dependent or not, what kind of data type it should be and the exact structural equation that it should follow. \code{node()} can only be used to define nodes at one specific point in time, while the \code{node\_td()} function should only be used to define time-varying nodes for discrete-time simulations.

	\item[{\code{add\_node()} or \code{DAG + node}}] allows the definition made by \code{node()} or \code{node\_td()} to be added to the \code{DAG} object.

	\item[{\code{plot.DAG()}}] directly plots a \code{DAG} object using the \pkg{ggplot2} library \citep{Wickham2016}.

	\item[{\code{summary.DAG()}}] may be used to print the underlying structural equations of all nodes in a \code{DAG} object.

	\item[{\code{sim\_from\_dag()}}] is one of the two core simulation functions. Given a fully specified \code{DAG} object that only includes nodes defined using \code{node()}, it randomly generates a \code{data.table} \citep{Barrett2024} according to the DGP specified by the \code{DAG}.

	\item[{\code{sim\_discrete\_time()}}] is the second core simulation function. Given a fully specified \code{DAG} object that includes one or multiple nodes added using the \code{node\_td()} function, and possibly one or multiple nodes added using the \code{node()} function, it randomly generates data according to the specified DGP using a discrete-time simulation approach. This is described in detail in Section~\ref{chap::discrete_time_sim}.

	\item[{\code{sim\_n\_datasets()}}] allows users to directly generate multiple datasets from a single \code{DAG}, possibly using parallel processing.

	\item[{\code{sim2data()}}] may be used to transform the output produced by the \code{sim\_discrete\_time()} function into either the \emph{wide}, \emph{long} or \emph{start-stop} format to make further usage of the generated data easier.
\end{description}

The package additionally includes multiple functions starting with \code{node\_}. These functions are used to generate data of different types and using different specifications. Usually they do not have to be called directly by the user. Instead they are specified as \code{type}s in the \code{node()} or \code{node\_td()} functions and only called internally. Some additional convenience functions are also included, but are not further discussed in this article. Instead we choose to focus on describing the core functionality in detail and refer the interested user to the official documentation for more information.

\subsection{Defining the DAG}

Regardless of which kind of data the user want to generate, it is always necessary to first define a \code{DAG} object which adequately describes the DGP the user wants to simulate. In most cases this should be done using the following steps:

\begin{enumerate}
	\item Initialize an empty \code{DAG} object using the \code{empty\_dag()} function.
	\item Define one or multiple nodes using the \code{node()} or \code{node\_td()} functions.
	\item Add these nodes to the \code{DAG} using the \code{+} syntax or the \code{add\_node()} function.
\end{enumerate}

The \code{empty\_dag()} function is very simple, as it does not have any arguments. It is merely used to setup the initial \code{DAG} object. The actual definition of the nodes should be done using either the \code{node()} function (node at a single point in time) or \code{node\_td()} function (node that varies over time), which have the following syntax:

\begin{Code}
node(name, type, parents=NULL, formula=NULL, ...)

node_td(name, type, parents=NULL, formula=NULL, ...)
\end{Code}

\par
The arguments are:

\begin{description}
	\item[{\code{name}}] A character string of the name of the node that should be generated, or a character vector including multiple names. If a character vector with more than one name is supplied, multiple independent nodes with the same definition will be added to the \code{DAG} object.

	\item[{\code{type}}] A character string specifying the type of the node or any suitable function that can be used to generate data for a node. Further details on supported node types are given in the Section~\ref{chap::node_types}.

	\item[{\code{parents}}] If the node is a child node this argument should contain the names of its' parents, unless a \code{formula} is supplied instead.

	\item[{\code{formula}}] An optional formula object that specifies an additive combination of coefficients and variables as required by generalized linear models for example. This argument may be used for most built-in node types and for user-defined functions. It allows inclusion of internally generated dummy variables (when using categorical variables as parents), interaction terms of arbitrarily high orders, cubic terms and arbitrary mixed model syntax for some node types.

	\item[{\code{...}}] Additional arguments passed to the function defined by the \code{type} argument.
\end{description}

For example, the simple DAG that we described earlier may be created using the following code:

\begin{Schunk}
\begin{Sinput}
R> library("simDAG")
R> dag <- empty_dag() +
+    node(c("A", "B"), type="rnorm", mean=0, sd=1) +
+    node("C", type="gaussian", formula=~-2 + A*0.3 + B*-2,
+         error=1)
\end{Sinput}
\end{Schunk}

First, an empty \code{DAG} object is initialized using the \code{empty\_dag()} function, to which nodes are added directly using the \code{+} syntax. Here, only the \code{node()} function is required, because all nodes only have to be defined for a single point in time, since this DAG is only supposed to describe crossectional data. Additionally, since both $A$ and $B$ have the same structural equation here, only one call to \code{node()} is needed to define both of these nodes. By setting \code{type="rnorm"} and leaving both the \code{parents} and the \code{formula} arguments at their default values, these nodes are specified as root nodes for which values will be generated using the \code{rnorm()} function with the additional arguments passed afterwards. Because $C$ is supposed to follow a linear regression model, \code{type="gaussian"} is used here and the structural equation is specified using the \code{formula} argument.
\par
The result is a \code{DAG} object. To re-create Figure~\ref{fig::example_dag}, users may use the associated S3 \code{plot()} method, which internally uses the \pkg{igraph} package \citep{Csardi2024} to put the nodes into position and the \pkg{ggplot2} package \citep{Wickham2016} and the \pkg{ggforce} \citep{Pedersen2022} for the actual plotting:

\begin{center}
\begin{Schunk}
\begin{Sinput}
R> library("igraph")
R> library("ggplot2")
R> library("ggforce")
R> plot(dag, layout="as_tree", node_size=0.15, node_fill="grey",
+       node_text_fontface="italic")
\end{Sinput}
\end{Schunk}
\includegraphics{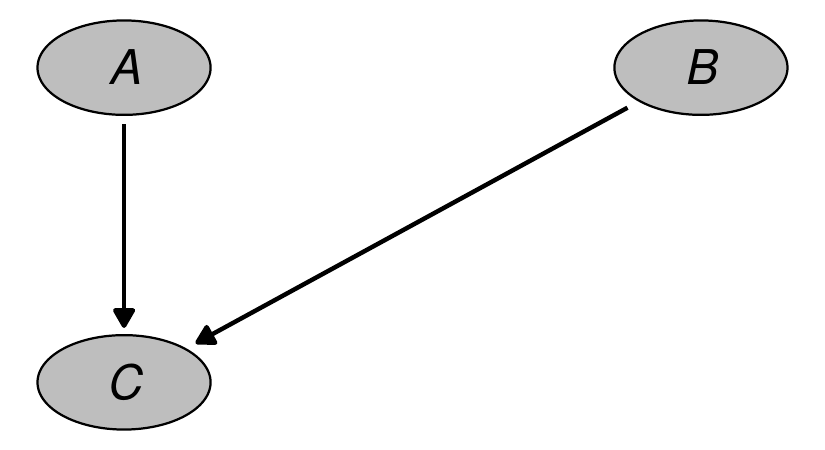}
\end{center}

The output is not exactly the same as Figure~\ref{fig::example_dag} because of space reasons, but it is very similar. Additionally, the underlying structural equations may be printed directly using the associated S3 \code{summary()} method:

\begin{Schunk}
\begin{Sinput}
R> summary(dag)
\end{Sinput}
\begin{Soutput}
A DAG object using the following structural equations:

A ~ N(0, 1)
B ~ N(0, 1)
C ~ N(-2 + A*0.3 + B*-2, 1)
\end{Soutput}
\end{Schunk}

Both of these methods may be useful to check whether the DAG is defined as intended by the user, or to formally describe the DGP in a publication.

\subsection{Supported node types} \label{chap::node_types}

Different \code{type}s of nodes are supported, depending on what kind of node is being specified. If the node is a root node, any function with an argument called \code{n} that specifies how many observations of some kind should be generated may be used. For example, the base \proglang{R} \code{runif()} function may be used as a type to generate a uniformly distributed node. Other popular choices are \code{rnorm()} for normally distributed nodes or \code{rgamma()} for a gamma distributed node. For convenience, the proposed package also includes implementations for fast Bernoulli trials (\code{rbernoulli()}), fast sampling from discrete probability distributions (\code{rcategorical()}) and a function to set a node to a constant value (\code{rconstant()}).
\par
If the node has one or more parent nodes, any function that can generate data based on these nodes and has the arguments \code{data} (containing the data generated up to this point) and \code{parents} (a vector with the names of the parent nodes) may be used as \code{type}. Multiple popular choices are directly implemented in this package. These include nodes based on generalized linear models and potentially more complex functions to sample from conditional distributions. Finally, the package also includes two specialized functions which may only be used for discrete-time simulations. These functions are able to generate binary or categorical time-varying covariates, as well as multiple forms of time-to-event data by repeatedly performing Bernoulli trials or multinomial trials over time. More details are given in Section~\ref{chap::discrete_time_sim}. A brief overview over all implemented node types is given in Table~\ref{tab::node_types}. Note that when using these node types, the user may either pass the respective function directly to the \code{type} argument, or use a string of the name without the \code{node\_} prefix.

\begin{table}[!htb]
	\centering
	\begin{tabular}{p{5.2cm} p{9.8cm}}
		\toprule
		\textbf{Node type} & \textbf{Description} \\
		\midrule
		\code{rbernoulli()} & Samples from a Bernoulli distribution. \\[0.1cm]
		\code{rcategorical()} & Samples from a discrete probability distribution. \\[0.1cm]
		\code{rconstant()} & Sets the node to a constant value. \\[0.1cm]
		\midrule
		\code{node\_gaussian()} & Generates a node based on a (mixed) linear regression model. \\[0.1cm]
		\code{node\_binomial()} & Generates a node based on a (mixed) logistic regression model. \\[0.1cm]
		\code{node\_conditional\_prob()} & Samples from a conditional discrete probability distribution. \\[0.1cm]
		\code{node\_conditional\_distr()} & Samples from different distributions conditional on values of other variables. \\[0.1cm]
		\code{node\_multinomial()} & Generates a node based on a multinomial regression model. \\[0.1cm]
		\code{node\_poisson()} & Generates a node based on a (mixed) Poisson regression model. \\[0.1cm]
		\code{node\_negative\_binomial()} & Generates a node based on a negative binomial regression model. \\[0.1cm]
		\code{node\_zeroinfl()} & Generates a node based on a zero-inflated Poisson or negative binomial regression model.\\[0.1cm]
		\code{node\_identity()} & Generates a node based on an \proglang{R} expression that includes previously generated nodes. \\[0.1cm]
		\code{node\_mixture()} & Generates a node as a mixture of other node types. \\[0.1cm]
		\code{node\_cox()} & Generates a node based on a Cox proportional hazards regression model, using the method of \citet{Bender2005}. \\[0.1cm]
		\code{node\_aftreg()}, \code{node\_ahreg()}, \code{node\_ehreg()}, \code{node\_poreg()}, \code{node\_ypreg()} & Generates a node based on various parametric survival models as implemented in \pkg{rsurv} \citep{Demarqui2024}. \\[0.1cm]
		\midrule
		\code{node\_time\_to\_event()} & A node based on repeated Bernoulli trials over time, intended to generate time-varying variables and time-to-event nodes. \\[0.1cm]
		\code{node\_competing\_events()} & A node based on repeated multinomial trials over time, intended to generate time-varying variables and time-to-event nodes.\\
		\bottomrule
	\end{tabular}
	\caption{A brief overview over all implemented node types. The first section contains functions that may only be used to generate root nodes, while the second section contains functions that may only be used to generate child nodes. Nodes in both of these sections may be used in both \code{node()} and \code{node\_td()} calls. The nodes mentioned in the third section are only meant to be used for discrete-time simulations.}
	\label{tab::node_types}
\end{table}

\FloatBarrier

\section{Simulating crossectional data}

In the following Section we will illustrate how to use the \pkg{simDAG} package to simulate more complex crossectional data. Instead of using a made up artificial example, we will do this by partially replicating a real Monte-Carlo simulation study by \citet{Denz2023}, published in the prestigious peer-reviewed journal \emph{Statistics in Medicine}. Because this package strictly focuses on the data generation step of Monte-Carlo studies and for reasons of brevity, we will not reproduce the entire simulation study. Instead we will only replicate the DGP used to generate the data for this study.
\par
\citet{Denz2023} recently performed a neutral comparison study of multiple different methods to estimate counterfactual survival curves from crossectional observational data. In this Monte-Carlo simulation study they wanted to investigate how different kinds of misspecifications of nuisance models, which are used in some methods to estimate the counterfactual survival curves, affect the estimates produced by the different methods. To do this, a DGP was required that includes multiple interrelated binary and continuous variables, as well as a right-censored time-to-event variable.
\par
In particular, the data sets they generated for each simulation run included six covariates, two of which were binary ($X_1, X_3$) and four of which were continuous ($X_2, X_4, X_5, X_6$). It additionally included a binary treatment variable ($Z$) and a right-censored time-to-event outcome ($T$). The two binary covariates $X_1$ and $X_3$ followed a simple Bernoulli distribution with a success probability of 0.5. $X_2$ was generated by a linear regression model, dependent on $X_3$. The two continuous covariates $X_4$ and $X_6$ were standard normally distributed, while $X_5$ was generated according to a linear regression model dependent on $X_6$. The treatment variable $Z$ followed a logistic regression model, dependent on $X_2$, $X_3$, $X_5$ and $X_6$, where $X_2$ was included as a squared term. Finally, the outcome $T$ was generated according to a Cox model, dependent on $X_1$, $X_2$, $X_4$, $X_5$ and $Z$, where $X_5$ was included as a squared term. Data for $T$ was generated using the method by \citep{Bender2005}, based on a Weibull distribution ($\lambda=2, \gamma=2.4$). The time until censoring was generated independently from a second Weibull distribution ($\lambda=1, \gamma=2$).
\par
This DGP was used because it includes two confounders ($X_2$, $X_5$) for the causal effect of $Z$ on $T$, which are correlated with other non-confounding variables. The inclusion of non-linear relationships allowed \citet{Denz2023} to investigate different kinds of misspecified models. More details on the DGP and the simulation study itself are given in the original manuscript. To replicate the DGP of this study, the following \code{DAG} definition may be used:

\begin{Schunk}
\begin{Sinput}
R> dag <- empty_dag() +
+    node(c("X1", "X3"), type="rbernoulli", p=0.5,
+         output="numeric") +
+    node(c("X4", "X6"), type="rnorm", mean=0, sd=1) +
+    node("X2", type="gaussian", formula=~0.3 + X3*0.1,
+         error=1) +
+    node("X5", type="gaussian", formula=~0.3 + X6*0.1,
+         error=1) +
+    node("Z", type="binomial", formula=~-1.2 + I(X2^2)*log(3)
+          + X3*log(1.5) + X5*log(1.5) + X6*log(2),
+         output="numeric") +
+    node("T", type="cox", formula=~X1*log(1.8) + X2*log(1.8) +
+          X4*log(1.8) + I(X5^2)*log(2.3) + Z*-1,
+         surv_dist="weibull", lambda=2, gamma=2.4,
+         cens_dist="rweibull",
+         cens_args=list(shape=1, scale=2))
\end{Sinput}
\end{Schunk}

As before, we can define the nodes $X_1$ and $X_3$ and the nodes $X_4$ and $X_5$ using a single call to \code{node()}, because they have the same definition. Since only directly supported regression models are required for this DGP, we were able to use the \code{formula} argument to directly type out all of the structural equations.
\par
Plotting this \code{DAG} using the \code{plot()} method produces the following output:

\begin{center}
\begin{Schunk}
\begin{Sinput}
R> plot(dag, node_size=0.3, node_fill="grey",
+       node_text_fontface="italic")
\end{Sinput}
\end{Schunk}
\includegraphics{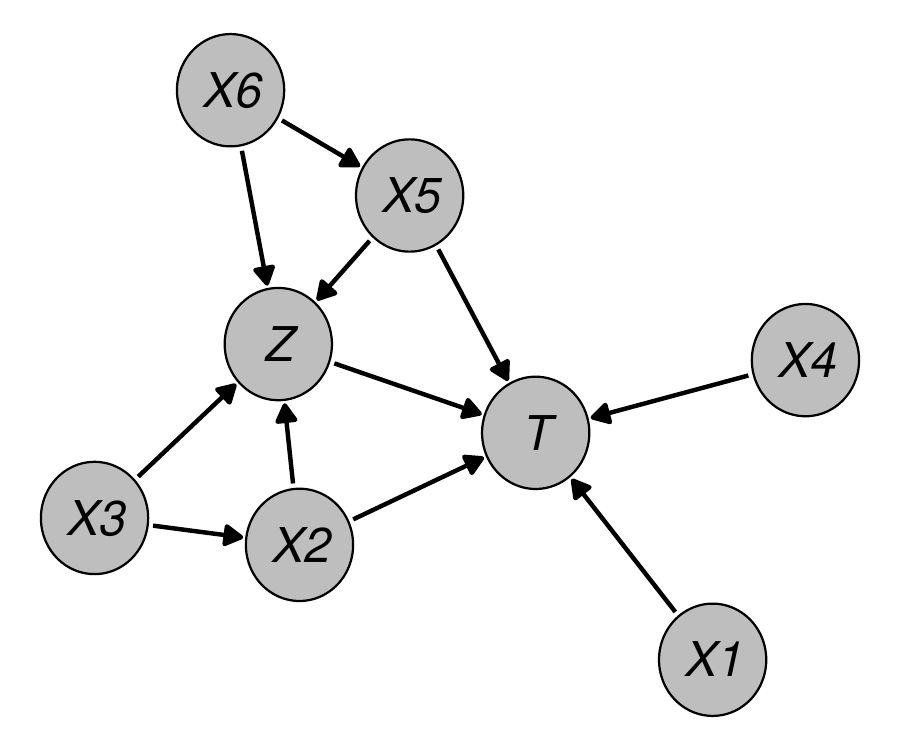}
\end{center}

\par
The underlying structural equations are:

\begin{Schunk}
\begin{Sinput}
R> summary(dag)
\end{Sinput}
\begin{Soutput}
A DAG object using the following structural equations:

  X1 ~ Bernoulli(0.5)
  X3 ~ Bernoulli(0.5)
  X4 ~ N(0, 1)
  X6 ~ N(0, 1)
  X2 ~ N(0.3 + X3*0.1, 1)
  X5 ~ N(0.3 + X6*0.1, 1)
   Z ~ Bernoulli(logit(-1.2 + I(X2^2)*log(3) + X3*log(1.5) +
                  X5*log(1.5) + X6*log(2)))
T[T] ~ (-(log(Unif(0, 1))/(2*exp(X1*log(1.8) + X2*log(1.8) +
         X4*log(1.8) + I(X5^2)*log(2.3) + Z*-1))))^(1/2.4)
T[C] ~ rweibull(shape=1, scale=2)
\end{Soutput}
\end{Schunk}

Finally, we can generate a single \code{data.table} from this \code{DAG} using the \code{sim\_from\_dag()} function. The first few rows of the generated data look like this:

\begin{Schunk}
\begin{Sinput}
R> library("data.table")
R> dat <- sim_from_dag(dag, n_sim=500)
R> head(round(dat, 3))
\end{Sinput}
\begin{Soutput}
      X1    X3     X4     X6     X2     X5     Z T_time T_status
   <num> <num>  <num>  <num>  <num>  <num> <num>  <num>    <num>
1:     0     0 -0.646  1.186 -0.628  1.395     1  0.318        0
2:     1     0  0.968 -2.368  0.101 -0.040     0  0.135        1
3:     0     0  0.125  2.012  0.662  0.694     1  0.758        1
4:     0     1 -0.142 -2.610 -0.826  0.588     1  1.100        1
5:     0     0 -0.273  0.632  1.462  0.007     1  0.890        1
6:     1     1 -0.198  0.379  0.693  2.416     1  0.058        1
\end{Soutput}
\end{Schunk}

Although only eight nodes were defined in the corresponding \code{DAG}, it includes nine columns. The reason for this is that nodes generated using \code{type="cox"} by default generate two columns: one for the observed time and one indicating whether the observation is right-censored or not. This is the standard data format for such time-to-event data and one of the reasons supplied node functions are allowed to output multiple columns at the same time. This of course also extends to custom user-defined node \code{type}s. When no censoring distribution is supplied, it would also be possible to return only the simulated time-to-event in a single column by setting \code{as_two_cols=FALSE} in the \code{node()} call for $T$.

\section{Simulating longitudinal data with few points in time} \label{chap::longitudinal_data_1}

For many applications the desired DGP may contain one or more variables that change over time. If that is the case, longitudinal data must be generated. There are two approaches to do this using the algorithm implemented in the proposed package: (1) defining one node per point in time of interest and (2) constructing one node definition for all points in time that should be considered simultaneously. In this Section the first approach will be illustrated by replicating the DGP that was used in the Monte-Carlo simulation study performed by \citet{Gruber2015}.
\par
In their study, \citet{Gruber2015} compared the efficiency of different methods of estimating inverse probability weights for marginal structural models. Such models are often used to estimate marginal causal effects of treatment-regimes from longitudinal observational data with time-varying variables \citep{Hernan2020}. The main aim was to quantify the benefits of using ensemble methods versus using traditional methods, such as logistic regression models, when estimating the required weights. Their DGP therefore required the inclusion of time-varying variables. Below we focus on the DGP theses authors used in their first simulation scenario.
\par
Their DGP consisted of four variables: a binary treatment variable $A$, a binary unmeasured baseline covariate $U$, a continuous confounder $L$ and a binary outcome $Y$. Since $U$ represents a baseline variable, it does not vary over time. All other variables, however, were generated at two distinct time points. $L$ and $A$ were generated for time $0$ and $1$, while the time-lagged outcome $Y$ was generated for times $1$ and $2$. For simplicity, $A$ was affected only by present and past $L$ and not by previous values of $A$ itself. $L$ on the other hand was caused by previous values of itself and by $A$ and $U$. Finally, $Y$ is only caused by $U$, meaning that neither the treatment nor the confounder have any actual effect on the outcome. To generate continuous child nodes they relied on linear regression models. For binary child nodes, logistic regression models were used. A more detailed description of the data generation process is given in the original article \citep{Gruber2015}. A \code{DAG} object to define this DGP in the proposed package is given below.

\begin{Schunk}
\begin{Sinput}
R> dag <- empty_dag() +
+    node("U", type="rbernoulli", p=0.5, output="numeric") +
+    node("L0", type="gaussian", formula=~0.1 + 0.6*U,
+         error=1) +
+    node("A0", type="binomial", formula=~-0.4 + 0.6*L0,
+         output="numeric") +
+    node("Y1", type="binomial", formula=~-3.5 + -0.9*U,
+         output="numeric") +
+    node("L1", type="gaussian", formula=~0.5 + 0.02*L0 +
+          0.5*U + 1.5*A0, error=1) +
+    node("A1", type="binomial", formula=~-0.4 + 0.02*L0 +
+          0.58*L1, output="numeric") +
+    node("Y2", type="binomial", formula=~-2.5 + 0.9*U,
+         output="numeric")
\end{Sinput}
\end{Schunk}

Shown graphically, the \code{DAG} looks like this:

\begin{center}
\begin{Schunk}
\begin{Sinput}
R> plot(dag, node_size=0.2, node_fill="grey",
+       node_text_fontface="italic", layout="in_circle")
\end{Sinput}
\end{Schunk}
\includegraphics{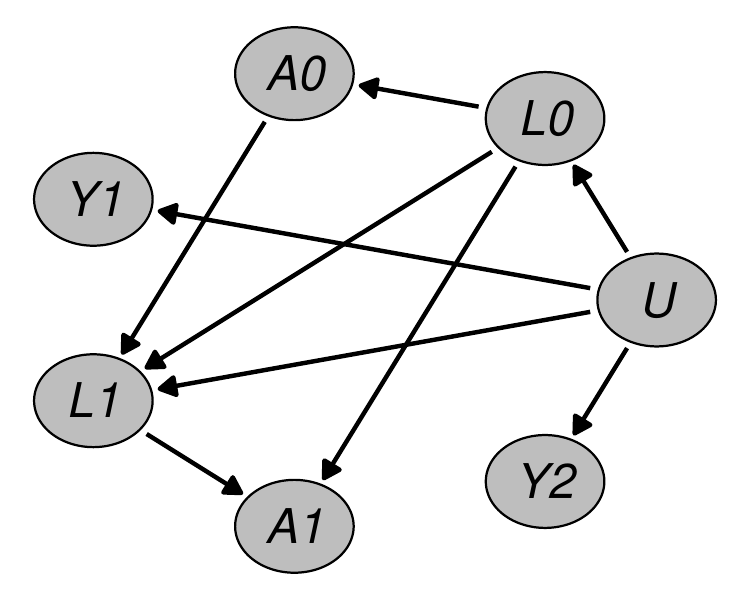}
\end{center}

The structural equations can again be printed using the \code{summary()} function:

\begin{Schunk}
\begin{Sinput}
R> summary(dag)
\end{Sinput}
\begin{Soutput}
A DAG object using the following structural equations:

 U ~ Bernoulli(0.5)
L0 ~ N(0.1 + 0.6*U, 1)
A0 ~ Bernoulli(logit(-0.4 + 0.6*L0))
Y1 ~ Bernoulli(logit(-3.5 + -0.9*U))
L1 ~ N(0.5 + 0.02*L0 + 0.5*U + 1.5*A0, 1)
A1 ~ Bernoulli(logit(-0.4 + 0.02*L0 + 0.58*L1))
Y2 ~ Bernoulli(logit(-2.5 + 0.9*U))
\end{Soutput}
\end{Schunk}

Finally, we can call the \code{sim\_from\_dag()} function on this \code{DAG} to generate some data:

\begin{Schunk}
\begin{Sinput}
R> dat <- sim_from_dag(dag, n_sim=1000)
R> head(dat)
\end{Sinput}
\begin{Soutput}
       U         L0    A0    Y1         L1    A1    Y2
   <num>      <num> <num> <num>      <num> <num> <num>
1:     0  1.3086008     0     0 -0.1642422     1     0
2:     1 -0.4978996     1     0  3.2424318     0     0
3:     1  0.5125773     1     0  2.4114821     1     0
4:     1  1.1811591     1     0  3.2285778     0     1
5:     1  1.3826790     1     0  1.7878382     1     0
6:     1 -0.3077301     0     0  1.1737307     0     0
\end{Soutput}
\end{Schunk}

Because the variables $A$, $L$ and $Y$ should be generated at different points in time, we have to include one node definition per variable per point in time to get an appropriate \code{DAG}. Apart from that the syntax is exactly the same as it was when generating crossectional data. More points in time could be added by simply adding more calls to \code{node()} to the \code{DAG} object, using appropriate regression models. The main advantage of this method is that it allows flexible changes of the DGP over time. However, the obvious shortcoming is that it does not easily extend to scenarios with many points in time. Although the authors only considered three time-varying variables and two distinct points in time here, the \code{DAG} definition is already quite cumbersome. For every further point in time, three new calls to \code{node()} would be necessary. If hundreds of points in time should be considered, using this method is essentially in-feasible. To circumvent these problems we describe a slightly different way of generating longitudinal data below.

\section{Simulating longitudinal data with many points in time} \label{chap::discrete_time_sim}

\subsection{Formal description}

Instead of defining each node for every point in time separately, it is often possible to define the structural equations in a generic time-dependent fashion, so that the same equation can be applied at each point in time, simplifying the workflow considerably. The result is the description of a specific stochastic process. For example, consider a very simple DAG with only one time-dependent node $Y$ at three points in time, $t \in \{0, 1, 2\}$:

\begin{figure}[!htb]
	\centering

\includegraphics{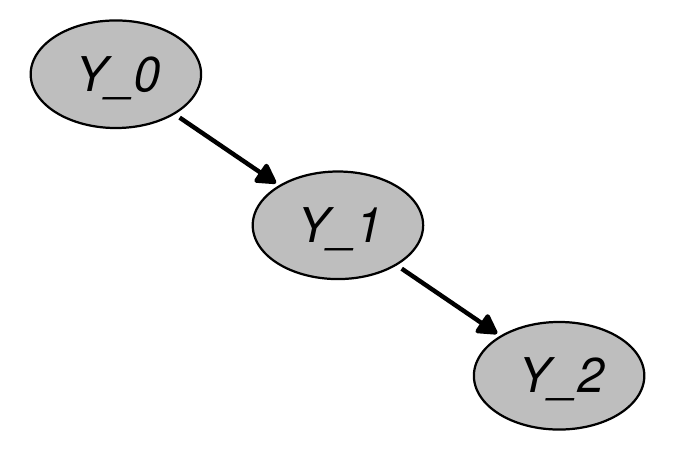}

	\caption{A simple DAG containing only one time-dependent node with three distinct points in time, that is only dependent on the last past value of itself.}
	\label{fig::dag_example2}
\end{figure}

In this DAG, $Y_t$ is only caused by values of itself at $t-1$. Suppose that $Y$ is a binary event indicator that is zero for everyone at $t = 0$. At every point in time $t$, $Y$ is set to 1 with a probability of 0.01. Once $Y$ is set to 1, it never changes back to 0. The following structural equation may be used to describe this DAG:

\begin{equation}
	Y_t \sim Bernoulli(P_Y(t)),
\end{equation}

where

\begin{equation} \label{eq::structural_eq_example2}
	P_Y(t) =
	\begin{cases}
		0 & \text{if} \quad t = 0 \\
		1 & \text{if} \quad Y_{t-1} = 1 \\
		0.01, & \text{otherwise}
	\end{cases}.
\end{equation}

The number of points in time could be increased by an arbitrary number and the same structural equation could still be used. Note that the time points may stand for any period of time, such as minutes, days or years. This is of course a trivial example, but this approach may also be used to define very complex DGPs, as illustrated below. For example, arbitrary dependencies on other variables measured at the same time or at any earlier time may be used when defining a node. To generate data from this type of DAG, the same algorithm as described in Section~\ref{chap::introduction} may be used. Since only discrete points in time are considered, this type of simulation has also been called \emph{discrete-time simulation} \citep{Tang2020} or \emph{dynamic microsimulation} \citep{Spooner2021} in the literature and is closely related to \emph{discrete-event simulation} \citep{Banks2014}.

\subsection{A simple example}

To generate data for the simple example considered above in the proposed package, we first have to define an appropriate \code{DAG} object as before. This can be done using the \code{node\_td()} function with \code{type="time\_to\_event"} as shown below.

\begin{Schunk}
\begin{Sinput}
R> dag <- empty_dag() +
+    node_td("Y", type="time_to_event", prob_fun=0.01,
+            event_duration=Inf)
\end{Sinput}
\end{Schunk}

By default, the value of node defined using \code{type="time\_to\_event"} is 0 for all individuals at $t = 0$. The \code{prob\_fun} argument defines the function that determines the occurrence probability at each point in time. It is set to 0.01 here, indicating that for all individuals and regardless of the value of $t$ the probability of experiencing the event is constant. Usually this argument will be passed an appropriate function to generate the occurrence probability for each individual at each point in time separately, but this is not necessary yet. By setting the \code{event\_duration} argument to \code{Inf}, we are indicating that the all events have an infinite duration, making the first event the final event for each person. In contrast to before, the \code{sim\_discrete\_time()} function now has to be used to generate data from this \code{DAG} object, because it contains a time-varying node:

\begin{Schunk}
\begin{Sinput}
R> sim <- sim_discrete_time(dag, n_sim=1000, max_t=80)
\end{Sinput}
\end{Schunk}

The \code{max\_t} argument specifies how many points in time should be simulated. Instead of just three points in time we consider 80 here. Contrary to the \code{sim\_from\_dag()} function, the \code{sim\_discrete\_time()} function does not return a single \code{data.table}. Instead it returns a \code{simDT} object, which usually has to be processed further using the \code{sim2data()} function to be useful. In this case, however, it is enough to simply extract the last state of the simulation to get all the information we need. This last simulation state is stored in the \code{\$data} parameter of the \code{simDT} object:

\begin{Schunk}
\begin{Sinput}
R> head(sim$data)
\end{Sinput}
\begin{Soutput}
     .id Y_event Y_time
   <int>  <lgcl>  <int>
1:     1   FALSE     NA
2:     2    TRUE     56
3:     3    TRUE      8
4:     4    TRUE     26
5:     5   FALSE     NA
6:     6    TRUE     58
\end{Soutput}
\end{Schunk}

As specified, the simulation contains only the variable $Y$, split into two columns. The first is called \code{Y\_event} and is a binary indicator of whether the individual is currently experiencing an event. The second column called \code{Y\_time} shows the time at which that event happened, or is set to \code{NA} if there is no event currently happening. Since every event is final, this is all information that was generated here. Individuals with a value of \code{NA} in \code{Y\_time} can be considered right-censored at $t = 80$. In this trivial example, it would be a lot easier to generate equivalent data by sampling from an appropriate parametric distributions. The following Section will illustrate the benefits of the approach using a more involved example.

\subsection{Simulating adverse events after Covid-19 vaccination} \label{chap::covid_example}

Suppose that we want to generate data for the Covid-19 pandemic, containing individual level information about Covid-19 vaccinations denoted by $A$ and the development of an acute myocarditis denoted by $Y$. Different individuals get vaccinated at different times, possible multiple times. Additionally, they might experience zero or multiple cases of myocarditis, also at different times. Both variables are therefore time-dependent binary variables, which are related to one another. If the target of interest is to estimate the effect of the vaccination on the time until the occurrence of a myocarditis, the vaccination may be considered a time-dependent variable and the myocarditis as a non-terminal, possibly recurrent, time-to-event outcome. This setup was of interest to \citet{Denz2023a}, who performed a Monte-Carlo simulation study to investigate the impact of linkage errors when estimating vaccine-safety from observational data. Below we will illustrate a simplified version of the DGP used therein.
\par
\begin{figure}[!htb]
	\centering
	\includegraphics[width=1\linewidth]{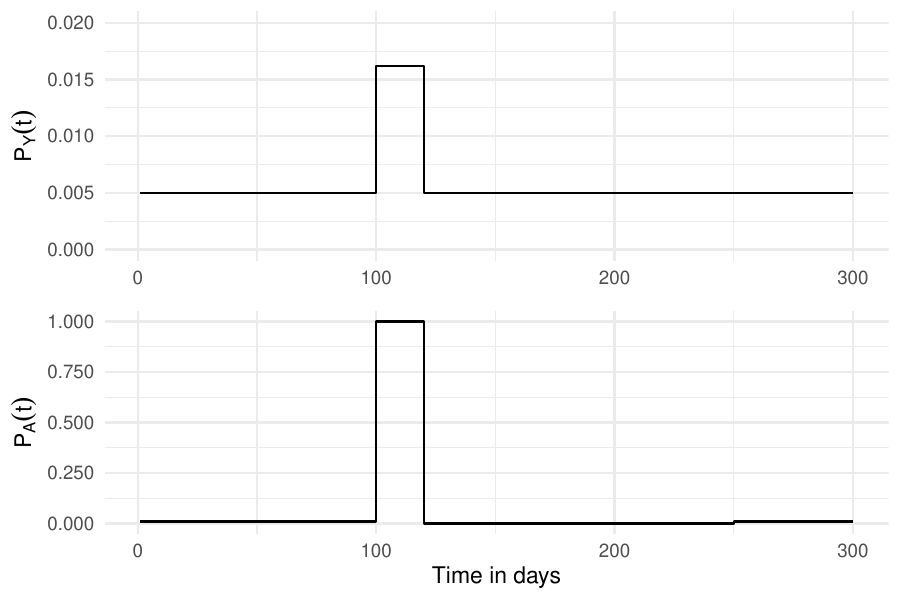}
	\caption{A simple graph showing $P_A(t)$ and $P_Y(t)$ for a fictional individual who got vaccinated once at $t = 100$, with $P_{A0} = 0.01$, $P_{Y0} = 0.005$, $d_{risk} = 20$ and $RR_A = 3.24$.}
	\label{fig::example_probs}
\end{figure}

For simplicity, we will make multiple simplifying assumptions that were not made in the original simulation study. First, we assume that the probability of being vaccinated and the base probability of developing a myocarditis are constant over time and equal for all individuals. The only risk factor for developing a myocarditis in this example is the Covid-19 vaccination itself. More precisely, the structural equation for the myocarditis node at $t$ is given by:

\begin{equation}
	Y_t \sim Bernoulli(P_{Y}(t)),
\end{equation}

with:

\begin{equation}
	P_{Y}(t) =
	\begin{cases}
		P_{Y0} \cdot RR_{A}, & \text{if } t \in  \left[T_{A}(t), T_{A}(t) + d_{risk}\right] \\
		P_{Y0}, & \text{otherwise}
	\end{cases},
\end{equation}

where $P_{Y0}$ denotes the base probability of developing a myocarditis, $T_{A}(t)$ denotes the time of the last performed vaccination and $d_{risk}$ defines the duration after the vaccination in which the risk of developing a myocarditis is elevated by $RR_{A}$. In this particular case, each $t$ will represent a single day. Similarly, the vaccination node can be described formally as:

\begin{equation} \label{eq::vacc_bernoulli}
	A_t \sim Bernoulli(P_{A}(t)),
\end{equation}

with:

\begin{equation} \label{eq::vacc_pt}
	P_{A}(t) =
	\begin{cases}
		1, & \text{if } t \in  \left[T_{A}(t), T_A(t) + 20 \right] \\
		0, & \text{if } t \in  \left[T_{A}(t) + 21, T_A(t) + 150\right] \\
		P_{A0}, & \text{otherwise}
	\end{cases},
\end{equation}

where $P_{A0}$ denotes the base probability of getting vaccinated. Figure~\ref{fig::example_probs} illustrates the result of applying these structural equations to a fictional person who gets vaccinated at $t = 100$.

The following code may be used to define this DGP:

\begin{Schunk}
\begin{Sinput}
R> prob_myoc <- function(data, P_0, RR_A) {
+    fifelse(data$A_event, P_0*RR_A, P_0)
+  }
R> dag <- empty_dag() +
+    node_td("A", type="time_to_event", prob_fun=0.01,
+            event_duration=20, immunity_duration=150) +
+    node_td("Y", type="time_to_event", prob_fun=prob_myoc,
+            parents=c("A_event"), P_0=0.005, RR_A=3.24)
\end{Sinput}
\end{Schunk}

First, we define a function that calculates $P_Y(t)$ at each simulated day for all individuals, called \code{prob\_myoc()}. This function simply checks whether the binary event indicator of the vaccination event, \code{"A\_event"}, is currently \code{TRUE} and multiplies the baseline probability of developing a myocarditis $P_{Y0}$ with the relative risk if this is the case. Otherwise it just returns the baseline probability directly. This function is then passed directly to the \code{prob\_fun} argument in the \code{node\_td()} call when defining the myocarditis node. For this to be a sensible strategy, we need to ensure that \code{"A\_event"} is only \code{TRUE} when a person is currently at elevated risk for a myocarditis, as defined in Equation~\ref{eq::vacc_pt}. In other words, \code{"A\_event"} should not actually be an indicator of whether someone just received a vaccination, but an indicator of whether the person is currently in the risk period of 20 days following the vaccination. We can achieve this by setting the \code{event\_duration} parameter in the node definition of the vaccination node to 20, meaning that the vaccination node will equal 1 for 20 days after a vaccination was performed. This argument is a direct feature of the \code{node_time_to_event()} function, which is called internally whenever \code{type="time_to_event"} is used in a time-dependent node.
\par
The base probability for the vaccination and for the myocarditis events are set to the arbitrary values of 0.01 and 0.005 respectively. $RR_{A}$ is set to 3.24, which is the value used in the actual simulation study by \citet{Denz2023a}. The \code{immunity\_duration} parameter used for the vaccination node additionally specifies that a person will not receive another vaccination in the first 150 days after a vaccination was performed. More specifically, these settings ensure that \code{"A\_event"} is set to \code{FALSE} for 130 days after the \code{event\_duration} of 20 days is over. This is another feature of \code{"time_to_event"} nodes. To run the simulation for two simulated years, the following code may be used:

\begin{Schunk}
\begin{Sinput}
R> sim <- sim_discrete_time(dag, n_sim=10000, max_t=365*2)
\end{Sinput}
\end{Schunk}

Since both of the included variables change over time and may have multiple events, it is not appropriate to just look at the last state of the simulation in this case. Instead we will have to use the \code{sim2data()} function to obtain a useful dataset. The following code may be used to get a dataset in the \emph{start-stop} format:

\begin{Schunk}
\begin{Sinput}
R> dat <- sim2data(sim, to="start_stop", target_event="Y",
+                  keep_only_first=TRUE, overlap=TRUE)
R> head(dat)
\end{Sinput}
\begin{Soutput}
     .id start  stop      A      Y
   <int> <int> <num> <lgcl> <lgcl>
1:     1     1    57  FALSE   TRUE
2:     2     1    21  FALSE  FALSE
3:     2    21    38   TRUE   TRUE
4:     3     1    50  FALSE  FALSE
5:     3    50    70   TRUE  FALSE
6:     3    70   272  FALSE  FALSE
\end{Soutput}
\end{Schunk}

In this format, there are multiple rows per person, corresponding to periods of time in which all variables stayed the same. These intervals are defined by the \code{start} column (time at which the period starts) and \code{stop} columns (time at which the period ends). The \code{overlap} argument specifies whether these intervals should be overlapping or not. By setting \code{target\_event="Y"}, the function treats the \code{Y} node as the outcome instead of as another time-dependent covariate. The resulting data is in exactly the format needed to fit standard time-to-event models, such as a Cox model with time-varying covariates \citep{Zhang2018a}. Using the \pkg{survival} package \citep{Therneau2024}, we can do this using the following code:

\begin{Schunk}
\begin{Sinput}
R> library("survival")
R> mod <- coxph(Surv(start, stop, Y) ~ A, data=dat)
R> summary(mod)
\end{Sinput}
\begin{Soutput}
Call:
coxph(formula = Surv(start, stop, Y) ~ A, data = dat)

  n= 25020, number of events= 9871 

         coef exp(coef) se(coef)     z Pr(>|z|)    
ATRUE 1.15906   3.18693  0.02377 48.76   <2e-16 ***
---
Signif. codes:  0 '***' 0.001 '**' 0.01 '*' 0.05 '.' 0.1 ' ' 1

      exp(coef) exp(-coef) lower .95 upper .95
ATRUE     3.187     0.3138     3.042     3.339

Concordance= 0.58  (se = 0.002 )
Likelihood ratio test= 1928  on 1 df,   p=<2e-16
Wald test            = 2378  on 1 df,   p=<2e-16
Score (logrank) test = 2648  on 1 df,   p=<2e-16
\end{Soutput}
\end{Schunk}

Because the DGP fulfills all assumptions of the Cox model and the sample size is very large, the resulting hazard ratio is close to the relative risk of 3.24 that were specified in the \code{DAG} object. Other types of models, such as discrete-time survival models, may require different data formats \citep{Tutz2016}. Because of this, the \code{sim2data()} function also supports the transformation into \emph{long-} and \emph{wide-} format data.
\par
Above we assumed that the presence of a myocarditis event has no effect on whether a person gets vaccinated or not. In reality, it is unlikely that a person who is currently experiencing a myocarditis would get a Covid-19 vaccine. We can include this into the DGP by modifying the \code{prob\_fun} argument when defining the vaccination node as shown below:

\begin{Schunk}
\begin{Sinput}
R> prob_vacc <- function(data, P_0) {
+    fifelse(data$Y_event, 0, P_0)
+  }
R> dag <- empty_dag() +
+    node_td("A", type="time_to_event", prob_fun=prob_vacc,
+            event_duration=20, immunity_duration=150, P_0=0.01) +
+    node_td("Y", type="time_to_event", prob_fun=prob_myoc,
+            parents=c("A_event"), P_0=0.005, RR_A=3.24,
+            event_duration=80)
\end{Sinput}
\end{Schunk}

In this extension, the probability of getting vaccinated is set to 0 whenever the respective person is currently experiencing a myocarditis event. By additionally setting the \code{event\_duration} parameter of the myocarditis node to the arbitrary value of 80, we are defining that in the 80 days after experiencing the event there will be no vaccination for that person. Data could be generated and transformed using the same code as before.

\subsection{Additionally supported features} \label{chap::dt_extentions}

The example shown above illustrates only a small fraction of the possible features that could be included into the DGP using the discrete-time simulation algorithm implemented in the proposed package. Below we name more features that could theoretically be added to the DGP. Examples for how to actually implement these extensions are given in appendix~\ref{appendix::dt_features}.
\par

\begin{description}
	\item[\emph{Time-Dependent Probabilities and Effects}] In the example above both the base probabilities and the relative risks were always set to a constant, regardless of the simulation time. This may also be changed by re-defining the \code{prob\_fun} argument to have a named argument called \code{sim\_time}. Internally, the current time of the simulation will then be passed directly to the \code{prob\_fun}, allowing users to define any time-depencies.

	\item[\emph{Non-Linear Effects}] Above, we only considered the simple scenario in which the probability of an event after the occurrence of another event follows a simple step function, e.g., it is set to $P_0$ in general and increased by a fixed factor in a fixed duration after an event. This may also be changed by using more complex definitions of \code{prob\_fun} that are not based only on the \code{\_event} column of the exposure, but directly on the time of occurrence.

	\item[\emph{Adding more Binary Time-Dependent Variables}] Only two time-dependent variables were considered in the example. It is of course possible to add as many further variables as desired by the user by simply adding more appropriate \code{node\_td()} calls to the \code{DAG} object. Both the \code{sim\_discrete\_time()} function and the \code{sim2data()} function will still work exactly as described earlier.

	\item[\emph{Adding Baseline Covariates}] It is of course also possible to define a \code{DAG} that contains both time-indepdenent and time-dependent variables at the same time. This can be achieved by adding nodes with the \code{node()} and \code{node\_td()} function. Data for the time-independent variables will then be generated first (at $t = 0$) and the time-dependent simulation will be performed as before, albeit possibly dependent on the time-independent variables.

	\item[\emph{Adding Categorical Time-Dependent Variables}] Sometimes it may not be appropriate to describe a time-varying variable using only two classes. For theses cases, the proposed package also includes the \code{"competing\_events"} node type, which is very similar to the \code{"time\_to\_event"} node type. The only difference is that instead of Bernoulli trials multinomial trials are performed at each point in time for all individuals, allowing the generation of mutually exclusive events.

	\item[\emph{Adding Arbitrarily Distributed Time-Dependent Variables}] As in the \\ \code{sim\_from\_dag()} function, any kind of function may be used as a \code{type} for time-dependent variables. By appropriately defining those, it is possible to generate count and continuous time-varying variables as well.
	
	\item[\emph{Adding Ordered Events}] Some events should only be possible once something else has already happened. For example, a person usually cannot receive a PhD before graduating high school. This type of data may also be generated, again by appropriately specifying the required probability functions.
\end{description}

All of the mentioned and shown features may of course be used in arbitrary combinations. Additionally, the list above is not meant to be exhaustive. It is merely supposed to highlight how versatile the implemented approach is.

\subsection{Computational considerations}

Although the discrete-time simulation procedure described above is incredibly flexible, it is also very computationally expensive theoretically. For example, when using nodes of type \code{"time\_to\_event"}. as shown above, the program has to re-calculate the event probabilities at each point in time, for every individual and for every node. Increasing either the number of nodes, the number of time points or the number of considered individuals therefore non-linearly increases the number of required computations. This fact, together with the lack of available appropriate software packages, is probably the main reason this type of simulation strategy is rarely used in Monte-Carlo simulation studies. Since Monte-Carlo studies often require thousands of replications with hundreds of different scenarios, even a runtime of a few seconds to generate a dataset may be too long to keep the simulation computationally feasible.
\par
The presented implementation was therefore designed explicitly to be as fast as possible. To ensure that it can be run on multiple processing cores in parallel, it is additionally designed to use as little random access memory (RAM) as possible. As all other code in this package, it exclusively uses the \pkg{data.table} package \citep{Barrett2024} to perform most required computations. The \pkg{data.table} package is arguably the best choice for doing data wrangling in the \proglang{R} ecosystem in terms of computational efficiency and has similar performance as corresponding software libraries in \proglang{Python} and \proglang{Julia} \citep{Chiou2023}. The proposed package additionally relies on a few tricks to keep the amount of memory used small. For example, when using nodes of type \code{"time\_to\_event"}, by default not every state of the simulation is saved. Only the times at which an event occurred are recorded. This information is then efficiently pieced together in the \code{sim2data()} function when creating the actual dataset.
\par
Although none of these code optimizations can entirely overcome the inherent computational complexity of the approach, they do ensure that the usage of this method is feasible to generate reasonably large data with many points in time on a regular office computer. For example, generating a dataset with 1000 individuals and 730 distinct points in time from the first \code{DAG} shown in Section~\ref{chap::covid_example} takes only 1.11 seconds on average on the authors personal computer, using a single processing core (Intel(R) Core(TM) i7-9700 CPU @ 3.00GHz, 32GB RAM). It is, however, also important to note that the runtime highly depends on how efficient the functions supplied to the \code{prob\_fun} arguments are written as well.
\par
Below we give some additional limited code to illustrate the increased runtime of the \code{sim\_discrete\_time()} function with increasing \code{n\_sim} and increasing \code{max\_t} values. As an example, we use the first DGP shown in Section~\ref{chap::covid_example} to generate some data for multiple different values of each argument. We include calls to the \code{sim2data()} function in the runtime calculations, because calling this function is necessary to obtain usable data and as such should be considered part of the DGP. The runtime is calculated using the \pkg{microbenchmark} package \citep{Mersmann2023}.

\begin{center}
\begin{Schunk}
\begin{Sinput}
R> library(microbenchmark)
R> set.seed(1234)
R> prob_myoc <- function(data, P_0, RR_A) {
+    fifelse(data$A_event, P_0*RR_A, P_0)
+  }
R> run_example <- function(n, max_t) {
+    dag <- empty_dag() +
+      node_td("A", type="time_to_event", prob_fun=0.01,
+              event_duration=20, immunity_duration=150) +
+      node_td("Y", type="time_to_event", prob_fun=prob_myoc,
+              parents=c("A_event"), P_0=0.005, RR_A=3.24)
+  
+    sim <- sim_discrete_time(dag, n_sim=n, max_t=max_t)
+  
+    dat <- sim2data(sim, to="start_stop", target_event="Y",
+                    keep_only_first=TRUE, overlap=TRUE)
+  }
R> n <- c(10, 100, 1000, 10000, 100000)
R> max_t <- c(10, 100, 1000, 10000)
R> params <- data.frame(max_t=rep(max_t, each=length(n)),
+                       n=rep(n), time=NA)
R> for (i in seq_len(nrow(params))) {
+    n_i <- params$n[i]
+    max_t_i <- params$max_t[i]
+  
+    bench <- microbenchmark(run_example(n=n_i, max_t=max_t_i),
+                            times=1)
+    params$time[i] <- mean(bench$time / 1000000000)
+  }
R> params <- within(params, {
+    max_t <- factor(max_t);
+  })
R> ggplot(params, aes(x=n, y=time, color=max_t)) +
+    geom_point() +
+    geom_line() +
+    theme_bw() +
+    labs(x="n_sim", y="Runtime in Seconds", color="max_t") +
+    scale_x_continuous(labels=scales::label_comma(),
+                       transform="log10") +
+    scale_y_log10()
\end{Sinput}
\end{Schunk}
\includegraphics{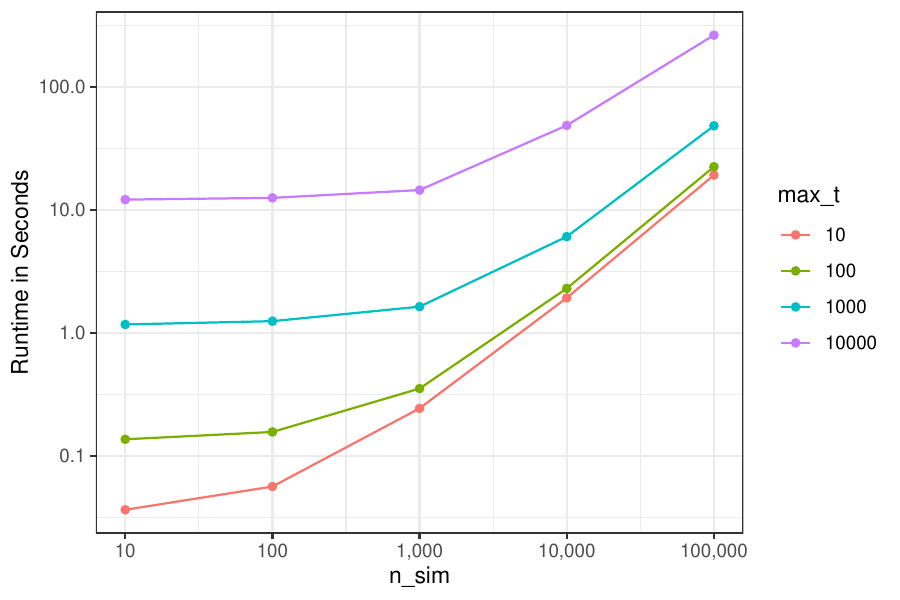}
\end{center}

The runtime increases more sharply with higher values of \code{max_t} than it does with higher values of \code{n_sim}, because each additional considered point in time internally translates to one more iteration in an \proglang{R} \code{for} loop, while each additional considered individual translates to one more row in the generated data, which is processed using optimized \code{data.table} code directly.

\section{Discussion}

In this article we presented the \pkg{simDAG} \proglang{R} package and illustrated how it can be used to generate various forms of artificial data that may be required for Monte-Carlo simulation studies. In particular, we showed how the package may be used to generate crossectional and longitudinal data with multiple variables of different data types, non-linear effects and arbitrary causal dependencies. We showed how similar the syntax is for very different kinds of DGP and how closely the required code resembles the actual underlying structural equations when using built-in node types. Because the package is based on defining a DAG to describe the DGP, it lends itself particularly well to simulation studies dealing with causal inference, but it is by no means limited to applications in this field, as shown in the second to last Section of the article.
\par
In addition to the main data generation functions, the package also includes multiple functions to facilitate the accurate description of the underlying DGP. Such descriptions are of great importance to facilitate both understanding and reproducibility of simulation studies, as emphasized in the literature \citep{Morris2019, Cheng2016}. Among these are the \code{plot.DAG()} function, that was used throughout the article to graphically display the defined \code{DAG} objects. While this function is useful on its own, some users may prefer to use one of the many other options to display DAGs in \proglang{R} \citep{Pitts2024}. To make this easier for users the package also contains the \code{dag2matrix()} function, which returns the underlying adjacency matrix of a \code{DAG} object and the \code{as.igraph.DAG()} function for direct integration with the \pkg{igraph} \proglang{R} package \citep{Csardi2024}. Additionally, the \code{summary.DAG()} method may be used to directly print the used structural equations, as shown throughout the article.
\par
The most distinguishing feature of the package is its capability of carrying out discrete-time simulations to generate longitudinal data with hundreds of points in time in a suitable amount of time. While other packages, such as the \pkg{simcausal} package \citep{Sofrygin2017} offer similar features to generate crossectional data, its' implementation for generation of longitudinal data is very different from the proposed package. In \pkg{simcausal} a new node is defined for each point in time internally. Although the user has direct access to each of these nodes (and therefore to each value at any point in time), the provided formula interface does not naturally support the definition of nodes with events that occur at some point in time and last for a certain amount of time. This can be done with little effort in the \pkg{simDAG} package using the provided \code{"time\_to\_event"} node type.
\par
This type of node can then be used to specify outcomes or to specify binary time-varying covariates, as illustrated in the main text where we used two such nodes to describe both the Covid-19 vaccination status and the associated adverse event. Using just this node type it is therefore possible to define DGPs including a time-to-event outcome (possibly with recurrent events) with multiple time-varying covariates. While there are multiple other methods to generate some forms of time-to-event data with time-varying covariates \citep{Hendry2013, Austin2012, Huang2020, Ngwa2022}, most of them require strict parametrizations or do not support the use of multiple, arbitrarily distributed time-varying variables. Additionally, neither of these methods allows the inclusion of recurrent events or competing events. None of these restrictions apply to the discrete-time simulation approach.
\par
However, the method also has two main disadvantages. First, it is a lot more computationally expensive than other methods, and secondly it does usually require the user to define appropriate functions to generate the time- and individual specific probabilities per node. Although the inherent computational complexity cannot be removed, it is alleviated in the implementation of this package through the use of optimized code and the \pkg{data.table} back-end \citep{Barrett2024}. As shown in the article, the presented implementation allows generation of large datasets in a reasonable amount of time. A computationally more efficient alternative not considered in this package would be \emph{discrete-event simulation} \citep{Banks2014}, in which the time until the next event is modeled directly instead of simulating the entire process over time. Performing such simulations is, however, usually a lot more demanding both conceptually and in terms of required software development \citep{Zhang2018}. The burden of specifying appropriate input to the \code{prob\_fun} argument in our approach is comparatively small, but it might still be a concern for some users. We hope that the many provided examples and explanations in both this article and the extensive documentation and multiple vignettes of the package will help users overcome this issue.
\par
To keep this article at a reasonable length, it was necessary to omit some implemented features of the \pkg{simDAG} package. One of these features is its capability to generate data for complex multi-level DGPs. By internally using code from the \pkg{simr} package \citep{Green2016}, it allows users to add arbitrarily complex random effects and random slopes to nodes of type \code{"gaussian"}, \code{"binomial"} and \code{"poisson"}. This can be done by directly adding the standard mixed model \proglang{R} syntax to the \code{formula} interface. Additionally, while the main focus of the package is on generating data from a user-defined DGP, the package offers limited support for generating data that mimics already existing data through the \code{dag\_from\_data()} function. This function only requires the user to specify the causal dependencies assumed to be present in the data and the type of each node and then directly extracts coefficients and intercepts from the available data by fitting various models to it. The returned DAG is then fully specified and can be used directly in a standard \code{sim\_from\_dag()} call to obtain data similar to the one supplied. Note that this function does not directly support all available node types. If the main goal of the user is to generate such synthetic mimicked datasets, using packages such as \pkg{simPop} \citep{Templ2017} might be preferable.
\par
Finally, we would like to note that the package is still under active development. We are currently working on multiple new features to make the package even more versatile for users. For example, future versions of the package are planned to support the definition of interventions on the DAG, much like the \pkg{simcausal} package \citep{Sofrygin2017}, which would make it even easier to generate data for causal inference based simulations, without having to re-define the DAG multiple times. We also plan to extend the internal library of available node types, by for example including node functions to simulate competing events data without the use of discrete-time simulation \citep{Morina2017, Haller2014}.


\section*{Computational details}

The results in this paper were obtained using
\proglang{R}~4.3.3 with the
\pkg{data.table}~1.15.2 package, the \pkg{survival}~3.5.8 package, the \pkg{igraph}~2.0.3 package, the \pkg{ggplot2}~3.5.0 package, the \pkg{microbenchmark}~1.4.10 package and the \pkg{simDAG}~0.3.1 package. \proglang{R} itself
and all packages used are available from the Comprehensive
\proglang{R} Archive Network (CRAN) at
\url{https://CRAN.R-project.org/}.

\section*{Acknowledgments}

We would like to thank Katharina Meiszl for her valuable input and for contributing some unit tests and input checks to the proposed package. We would also like to thank the anonymous reviewers, whose feedback and suggestions greatly improved the manuscript and the package itself.

\bibliography{disc_time_sim}

\begin{thebibliography}{58}
\newcommand{\enquote}[1]{``#1''}
\providecommand{\natexlab}[1]{#1}
\providecommand{\url}[1]{\texttt{#1}}
\providecommand{\urlprefix}{URL }
\expandafter\ifx\csname urlstyle\endcsname\relax
  \providecommand{\doi}[1]{doi:\discretionary{}{}{}#1}\else
  \providecommand{\doi}{doi:\discretionary{}{}{}\begingroup
  \urlstyle{rm}\Url}\fi
\providecommand{\eprint}[2][]{\url{#2}}

\bibitem[{Alfons \emph{et~al.}(2010)Alfons, Templ, and Filzmoser}]{Alfons2010}
Alfons A, Templ M, Filzmoser P (2010).
\newblock \enquote{An Object-Oriented Framework for Statistical Simulation: The
  \proglang{R} Package \pkg{simFrame}.}
\newblock \emph{Journal of Statistical Software}, \textbf{37}(3), 1--36.
\newblock \doi{10.18637/jss.v037.i03}.

\bibitem[{Andrews and Kummerfeld(2024)}]{Andrews2024}
Andrews B, Kummerfeld E (2024).
\newblock \enquote{Better Simulations for Validating Causal Discovery with the
  DAG-Adaptation of the Onion Method.}
\newblock arXiv:2405.13100v1.

\bibitem[{Arnold \emph{et~al.}(2011)Arnold, Hogan, {Colford Jr.}, and
  Hubbard}]{Arnold2011}
Arnold BF, Hogan DR, {Colford Jr} JM, Hubbard AE (2011).
\newblock \enquote{Simulation Methods to Estimate Design Power: An Overview for
  Applied Research.}
\newblock \emph{BMC Medical Research Methodology}, \textbf{11}(94), 1--10.
\newblock \doi{10.1186/1471-2288-11-94}.

\bibitem[{Asparouhov and Muthén(2020)}]{Asparouhov2020}
Asparouhov T, Muthén B (2020).
\newblock \enquote{Comparison of Models for the Analysis of Intensive
  Longitudinal Data.}
\newblock \emph{Structural Equation Modeling: A Multidisciplinary Journal},
  \textbf{27}(2), 275--297.
\newblock \doi{10.1080/10705511.2019.1626733}.

\bibitem[{Austin(2012)}]{Austin2012}
Austin PC (2012).
\newblock \enquote{Generating Survival Times to Simulate Cox Proportional
  Hazards Models with Time-Varying Covariates.}
\newblock \emph{Statistics in Medicine}, \textbf{31}(29), 3946--3958.
\newblock \doi{10.1002/sim.5452}.

\bibitem[{Banks \emph{et~al.}(2014)Banks, {Carson II}, Nelson, and
  Nicol}]{Banks2014}
Banks J, {Carson II} JS, Nelson BL, Nicol DM (2014).
\newblock \emph{Discrete-Event System Simulation}, volume~5.
\newblock Pearson Education Limited, Edinburgh Gate.

\bibitem[{Barrett \emph{et~al.}(2024)Barrett, Dowle, Srinivasan, Gorecki,
  Chirico, and Hocking}]{Barrett2024}
Barrett T, Dowle M, Srinivasan A, Gorecki J, Chirico M, Hocking T (2024).
\newblock \emph{\pkg{data.table}: Extension of `data.frame`}.
\newblock \proglang{R} package version 1.15.2,
  \urlprefix\url{https://CRAN.R-project.org/package=data.table}.

\bibitem[{Bender \emph{et~al.}(2005)Bender, Augustin, and
  Blettner}]{Bender2005}
Bender R, Augustin T, Blettner M (2005).
\newblock \enquote{Generating Survival Times to Simulate Cox Proportional
  Hazards Models.}
\newblock \emph{Statistics in Medicine}, \textbf{24}(11), 1713--1723.
\newblock \doi{10.1002/sim.2059}.

\bibitem[{Brilleman \emph{et~al.}(2021)Brilleman, Wolfe, Moreno-Betancur, and
  Crowther}]{Brilleman2021}
Brilleman SL, Wolfe R, Moreno-Betancur M, Crowther MJ (2021).
\newblock \enquote{Simulating Survival Data Using the \pkg{simsurv}
  \proglang{R} Package.}
\newblock \emph{Journal of Statistical Software}, \textbf{97}(3).
\newblock \doi{10.18637/jss.v097.i03}.

\bibitem[{Byeon and Lee(2023)}]{Byeon2023}
Byeon S, Lee W (2023).
\newblock \enquote{Directed Acyclic Graphs for Clinical Research: A Tutorial.}
\newblock \emph{Journal of Minimally Invasive Surgery}, \textbf{26}(3),
  97--107.
\newblock \doi{10.7602/jmis.2023.26.3.97}.

\bibitem[{Carsey and Harden(2014)}]{Carsey2014}
Carsey TM, Harden JJ (2014).
\newblock \emph{Monte Carlo Simulation and Resampling Methods for Social
  Science}.
\newblock SAGE Publications, Thousand Oaks.

\bibitem[{Cheng \emph{et~al.}(2016)Cheng, Kessler, Mackinnon, Chang, Nadkarni,
  Hunt, Duval-Arnould, Lin, Cook, Pusic, Hui, Moher, Egger, and
  Auerbach}]{Cheng2016}
Cheng A, Kessler D, Mackinnon R, Chang TP, Nadkarni VM, Hunt EA, Duval-Arnould
  J, Lin Y, Cook DA, Pusic M, Hui J, Moher D, Egger M, Auerbach M (2016).
\newblock \enquote{Reporting Guidelines for Health Care Simulation Research:
  Extensions to the CONSORT and STROBE Statements.}
\newblock \emph{Advances in Simulation}, \textbf{1}(25), 1--13.
\newblock \doi{10.1186/s41077-016-0025-y}.

\bibitem[{Chiou \emph{et~al.}(2023)Chiou, Xu, and Huang}]{Chiou2023}
Chiou SH, Xu G, Huang JYCY (2023).
\newblock \enquote{Regression Modeling for Recurrent Events Possibly with an
  Informative Terminal Event Using \proglang{R} Package \pkg{reReg}.}
\newblock \emph{Journal of Statistical Software}, \textbf{105}(5), 1--34.
\newblock \doi{10.18637/jss.v105.i05}.

\bibitem[{Csárdi \emph{et~al.}(2024)Csárdi, Nepusz, Traag, Horvát, Zanini,
  Noom, and Müller}]{Csardi2024}
Csárdi G, Nepusz T, Traag V, Horvát S, Zanini F, Noom D, Müller K (2024).
\newblock \emph{\pkg{igraph}: Network Analysis and Visualization in
  \proglang{R}}.
\newblock \doi{10.5281/zenodo.7682609}.
\newblock \proglang{R} package version 2.0.3,
  \urlprefix\url{https://CRAN.R-project.org/package=igraph}.

\bibitem[{Demarqui(2024)}]{Demarqui2024}
Demarqui FN (2024).
\newblock \enquote{Survival Data Simulation with the \proglang{R} Package
  \pkg{rsurv}.}
\newblock arXiv:2406.01750v1.

\bibitem[{Denz \emph{et~al.}(2023{\natexlab{a}})Denz, Klaaßen-Mielke, and
  Timmesfeld}]{Denz2023}
Denz R, Klaaßen-Mielke R, Timmesfeld N (2023{\natexlab{a}}).
\newblock \enquote{A Comparison of Different Methods to Adjust Survival Curves
  for Confounders.}
\newblock \emph{Statistics in Medicine}, \textbf{42}(10), 1461--1479.
\newblock \doi{10.1002/sim.9681}.

\bibitem[{Denz \emph{et~al.}(2023{\natexlab{b}})Denz, Meiszl, Ihle, Oberle,
  Drechsel-Bäuerle, Scholz, Meyer, and Timmesfeld}]{Denz2023a}
Denz R, Meiszl K, Ihle P, Oberle D, Drechsel-Bäuerle U, Scholz K, Meyer I,
  Timmesfeld N (2023{\natexlab{b}}).
\newblock \enquote{Impact of Record-Linkage Errors in Covid-19 Vaccine-Safety
  Analyses using German Health-Care Data: A Simulation Study.}
\newblock arXiv:2310.15016v1.

\bibitem[{Fox \emph{et~al.}(2022)Fox, Nianogo, Rudolph, and Howe}]{Fox2022}
Fox MP, Nianogo R, Rudolph JE, Howe CJ (2022).
\newblock \enquote{Illustrating How to Simulate Data from Directed Acyclic
  Graphs to Understand Epidemiologic Concepts.}
\newblock \emph{American Journal of Epidemiology}, \textbf{191}(7), 1300--1306.
\newblock \doi{10.1093/aje/kwac041}.

\bibitem[{Goldfeld and Wujciak-Jens(2020)}]{Goldfeld2020}
Goldfeld K, Wujciak-Jens J (2020).
\newblock \enquote{\pkg{simstudy}: Illuminating Research Methods through Data
  Generation.}
\newblock \emph{The Journal of Open Source Software}, \textbf{5}(54), 1--4.
\newblock \doi{10.21105/joss.02763}.

\bibitem[{Green and MacLeod(2016)}]{Green2016}
Green P, MacLeod CJ (2016).
\newblock \enquote{\pkg{simr}: An R Package for Power Analysis of Generalized
  Linear Mixed Models by Simulation.}
\newblock \emph{Methods in Ecology and Evolution}, \textbf{7}(4), 493--498.
\newblock \doi{10.1111/2041-210X.12504}.

\bibitem[{Gruber \emph{et~al.}(2015)Gruber, Logan, Jarrín, Monge, and
  Hernán}]{Gruber2015}
Gruber S, Logan RW, Jarrín I, Monge S, Hernán MA (2015).
\newblock \enquote{Ensemble Learning of Inverse Probability Weights for
  Marginal Structural Modeling in Large Observational Datasets.}
\newblock \emph{Statistics in Medicine}, \textbf{34}(1), 106--117.
\newblock \doi{10.1002/sim.6322}.

\bibitem[{Hajj \emph{et~al.}(2023)Hajj, Pensar, and Sandve}]{Hajj2023}
Hajj GSA, Pensar J, Sandve GK (2023).
\newblock \enquote{\pkg{DagSim}: Combining DAG-Based Model Structure with
  Unconstrained Data Types and Relations for Flexible, Transparent, and
  Modularized Data Simulation.}
\newblock \emph{PLoS One}, \textbf{18}(4).
\newblock \doi{10.1371/journal.pone.0284443}.

\bibitem[{Haller and Ulm(2014)}]{Haller2014}
Haller B, Ulm K (2014).
\newblock \enquote{Flexible Simulation of Competing Risks Data Following
  Prespecified Subdistribution Hazards.}
\newblock \emph{Journal of Statistical Computation and Simulation},
  \textbf{84}(12), 2557--2576.

\bibitem[{Hendry(2013)}]{Hendry2013}
Hendry DJ (2013).
\newblock \enquote{Data Generation for the Cox Proportional Hazards Model with
  Time-Dependent Covariates: A Method for Medical Researchers.}
\newblock \emph{Statistics in Medicine}, \textbf{33}(3), 436--454.
\newblock \doi{10.1002/sim.5945}.

\bibitem[{Hernán and Robins(2020)}]{Hernan2020}
Hernán MA, Robins JM (2020).
\newblock \emph{Causal Inference: What If}.
\newblock CRC Press.

\bibitem[{Huang \emph{et~al.}(2020)Huang, Zhang, Zhang, and
  Gilbert}]{Huang2020}
Huang Y, Zhang Y, Zhang Z, Gilbert PB (2020).
\newblock \enquote{Generating Survival Times Using Cox Proportional Hazards
  Models with Cyclic and Piecewise Time-Varying Covariates.}
\newblock \emph{Statistics in Biosciences}, \textbf{12}, 324--339.
\newblock \doi{10.1007/s12561-020-09266-3}.

\bibitem[{Imbens(2020)}]{Imbens2020}
Imbens GW (2020).
\newblock \enquote{Potential Outcome and Directed Acyclic Graph Approaches to
  Causality: Relevance for Empirical Practice in Economics.}
\newblock \emph{Journal of Economic Literature}, \textbf{58}(4), 1129--1179.
\newblock \doi{10.1257/jel.20191597}.

\bibitem[{Jorgensen \emph{et~al.}(2022)Jorgensen, Pornprasertmanit, Schoemann,
  and Rosseel}]{Jorgensen2022}
Jorgensen TD, Pornprasertmanit S, Schoemann AM, Rosseel Y (2022).
\newblock \emph{\pkg{semTools}: {U}seful tools for structural equation
  modeling}.
\newblock \proglang{R} package version 0.5-6,
  \urlprefix\url{https://CRAN.R-project.org/package=semTools}.

\bibitem[{Kahn(1962)}]{Kahn1962}
Kahn AB (1962).
\newblock \enquote{Topological Sorting of Large Networks.}
\newblock \emph{Communications of the ACM}, \textbf{5}(11), 558--562.
\newblock \doi{10.1145/368996.369025}.

\bibitem[{Kimko and Duffull(2002)}]{Kimko2002}
Kimko HC, Duffull SB (eds.) (2002).
\newblock \emph{Simulation for Designing Clinical Trials: A
  Pharmacokinetic-Pharmacodynamic Modeling Perspective}.
\newblock Marcel Dekker Inc., New York.

\bibitem[{Kline(2023)}]{Kline2023}
Kline RB (2023).
\newblock \emph{Principles and Practice of Structural Equation Modeling}.
\newblock 5. The Guilford Press, New York.

\bibitem[{Mersmann(2023)}]{Mersmann2023}
Mersmann O (2023).
\newblock \emph{\pkg{microbenchmark}: Accurate Timing Functions}.
\newblock \proglang{R} package version 1.4.10,
  \urlprefix\url{https://CRAN.R-project.org/package=microbenchmark}.

\bibitem[{{Mori{\~{n}a}} and Navarro(2017)}]{Morina2017}
{Mori{\~{n}a}} D, Navarro A (2017).
\newblock \enquote{Competing Risks Simulation with the \pkg{survsim}
  \proglang{R} Package.}
\newblock \emph{Communications in Statistics: Simulation and Computation},
  \textbf{46}(7), 5712--5722.
\newblock \doi{10.18637/jss.v059.i02}.

\bibitem[{Morris \emph{et~al.}(2019)Morris, White, and Crowther}]{Morris2019}
Morris TP, White IR, Crowther MJ (2019).
\newblock \enquote{Using Simulation Studies to Evaluate Statistical Methods.}
\newblock \emph{Statistics in Medicine}, \textbf{38}(11), 2074--2102.
\newblock \doi{10.1002/sim.8086}.

\bibitem[{Nance \emph{et~al.}(2024)Nance, Petersen, {van der Laan}, and
  Balzer}]{Nance2024}
Nance N, Petersen ML, {van der Laan} MJ, Balzer LB (2024).
\newblock \enquote{The Causal Roadmap and Simulations to Improve the Rigor and
  Reproducibility of Real-Data Applications.}
\newblock \emph{Epidemiology}, \textbf{35}(6), 791--800.
\newblock \doi{10.1097/EDE.0000000000001773}.

\bibitem[{Ngwa \emph{et~al.}(2022)Ngwa, Cabral, Cheng, Gagnon, LaValley, and
  Cupples}]{Ngwa2022}
Ngwa JS, Cabral HJ, Cheng DM, Gagnon DR, LaValley MP, Cupples LA (2022).
\newblock \enquote{Generating Survival Times with Time-Varying Covariates Using
  the Lambert W Function.}
\newblock \emph{Communications in Statistics: Simulation and Computation},
  \textbf{51}(1), 135--153.
\newblock \doi{10.1080/03610918.2019.1648822}.

\bibitem[{Pearl(1995)}]{Pearl1995}
Pearl J (1995).
\newblock \enquote{Causal Diagrams for Empirical Research.}
\newblock \emph{Biometrika}, \textbf{82}(4), 669--688.
\newblock \doi{10.2307/2337329}.

\bibitem[{Pearl(2009)}]{Pearl2009}
Pearl J (2009).
\newblock \emph{Causality: Models, Reasoning and Inference}.
\newblock 2 edition. Cambridge University Press, Cambridge.

\bibitem[{Pedersen(2022)}]{Pedersen2022}
Pedersen TL (2022).
\newblock \emph{\pkg{ggforce}: Accelerating \pkg{ggplot2}}.
\newblock \proglang{R} package version 0.4.1,
  \urlprefix\url{https://CRAN.R-project.org/package=ggforce}.

\bibitem[{Pitts and Fowler(2024)}]{Pitts2024}
Pitts AJ, Fowler CR (2024).
\newblock \enquote{Comparison of Open-Source Software for Producing Directed
  Acyclic Graphs.}
\newblock \emph{Journal of Causal Inference}, \textbf{12}(1), 1--10.
\newblock \doi{10.1515/jci-2023-0031}.

\bibitem[{Pornprasertmanit \emph{et~al.}(2021)Pornprasertmanit, Miller,
  Schoemann, and Jorgensen}]{Pornprasertmanit2021}
Pornprasertmanit S, Miller P, Schoemann A, Jorgensen TD (2021).
\newblock \emph{\pkg{simsem}: SIMulated Structural Equation Modeling}.
\newblock \proglang{R} package version 0.5-16,
  \urlprefix\url{https://CRAN.R-project.org/package=simsem}.

\bibitem[{{\proglang{R} Core Team}(2024)}]{RCT2024}
{\proglang{R} Core Team} (2024).
\newblock \enquote{\proglang{R}: A Language and Environment for Statistical
  Computing.}
\newblock \proglang{R} Foundation for Statistical Computing, Vienna, Austria.
\newblock \urlprefix\url{https://www.r-project.org/}.

\bibitem[{Reisach \emph{et~al.}(2021)Reisach, Seiler, and
  Weichwald}]{Reisach2021}
Reisach AG, Seiler C, Weichwald S (2021).
\newblock \enquote{Beware of the Simulated DAG! Causal Discovery Benchmarks May
  Be Easy To Game.}
\newblock In M~Ranzato, A~Beygelzimer, YN~Dauphin, P~Liang, JW~Vaughan (eds.),
  \emph{Advances in Neural Information Processing Systems 34 (NeurIPS 2021)}.

\bibitem[{Rosseel(2012)}]{Rosseel2012}
Rosseel Y (2012).
\newblock \enquote{\pkg{lavaan}: An \proglang{R} Package for Structural
  Equation Modeling.}
\newblock \emph{Journal of Statistical Software}, \textbf{48}(2), 1--36.
\newblock \doi{10.18637/jss.v048.i02}.

\bibitem[{Sigal and Chalmers(2016)}]{Sigal2016}
Sigal MJ, Chalmers RP (2016).
\newblock \enquote{Play it Again: Teaching Statistics with Monte Carlo
  Simulation.}
\newblock \emph{Journal of Statistics Education}, \textbf{24}(3), 136--156.
\newblock \doi{10.1080/10691898.2016.1246953}.

\bibitem[{Sofrygin \emph{et~al.}(2017)Sofrygin, {van der Laan}, and
  Neugebauer}]{Sofrygin2017}
Sofrygin O, {van der Laan} MJ, Neugebauer R (2017).
\newblock \enquote{\pkg{simcausal} \proglang{R} Package: Conducting Transparent
  and Reproducible Simulation Studies of Causal Effect Estimation with Complex
  Longitudinal Data.}
\newblock \emph{Journal of Statistical Software}, \textbf{81}(2), 1--47.
\newblock \doi{10.18637/jss.v081.i02}.

\bibitem[{Spirtes \emph{et~al.}(2000)Spirtes, Glymour, and
  Scheines}]{Spirtes2000}
Spirtes P, Glymour C, Scheines R (2000).
\newblock \emph{Causation, Prediction, and Search}.
\newblock 2. MIT Press, Cambridge.

\bibitem[{Spooner \emph{et~al.}(2021)Spooner, Abrams, Morrissey, Shaddick,
  Batty, Milton, Dennett, Lomax, Malleson, Nelissen, Coleman, Nur, Jin, Greig,
  Shenton, and Birkin}]{Spooner2021}
Spooner F, Abrams JF, Morrissey K, Shaddick G, Batty M, Milton R, Dennett A,
  Lomax N, Malleson N, Nelissen N, Coleman A, Nur J, Jin Y, Greig R, Shenton C,
  Birkin M (2021).
\newblock \enquote{A Dynamic Microsimulation Model for Epidemics.}
\newblock \emph{Social Science \& Medicine}, \textbf{291}(114461).
\newblock \doi{10.1016/j.socscimed.2021.114461}.

\bibitem[{Tang \emph{et~al.}(2020)Tang, Leu, and Abbass}]{Tang2020}
Tang J, Leu G, Abbass HA (2020).
\newblock \emph{Simulation and Computational Red Teaming for Problem Solving}.
\newblock John Wiley \& Sons, Hoboke.

\bibitem[{Templ \emph{et~al.}(2017)Templ, Meindl, Kowarik, and
  Dupriez}]{Templ2017}
Templ M, Meindl B, Kowarik A, Dupriez O (2017).
\newblock \enquote{Simulation of Synthetic Complex Data: The \proglang{R}
  Package \pkg{simPop}.}
\newblock \emph{Journal of Statistical Software}, \textbf{79}(10), 1--38.
\newblock \doi{10.18637/jss.v079.i10}.

\bibitem[{Textor \emph{et~al.}(2016)Textor, {van der Zander}, Gilthorpe,
  Liśkiewicz, and Ellison}]{Textor2016}
Textor J, {van der Zander} B, Gilthorpe MS, Liśkiewicz M, Ellison GTH (2016).
\newblock \enquote{Robust Causal Inference using Directed Acyclic Graphs: The
  \proglang{R} Package \pkg{daggity}.}
\newblock \emph{International Journal of Epidemiology}, \textbf{45}(6),
  1887--1894.
\newblock \doi{10.1093/ije/dyw341}.

\bibitem[{Therneau(2024)}]{Therneau2024}
Therneau TM (2024).
\newblock \emph{A Package for Survival Analysis in \proglang{R}}.
\newblock \proglang{R} package version 3.5-8,
  \urlprefix\url{https://CRAN.R-project.org/package=survival}.

\bibitem[{Tutz and Schmid(2016)}]{Tutz2016}
Tutz G, Schmid M (2016).
\newblock \emph{Modeling Discrete Time-to-Event Data}.
\newblock Springer International Publishing Switzerland, Cham.

\bibitem[{Wang \emph{et~al.}(2022)Wang, Fu, Han, and Yan}]{Wang2022}
Wang W, Fu H, Han S, Yan J (2022).
\newblock \emph{\pkg{reda}: Recurrent Event Data Analysis}.
\newblock \proglang{R} package version 0.5.4.

\bibitem[{Wickham(2016)}]{Wickham2016}
Wickham H (2016).
\newblock \emph{\pkg{ggplot2}: Elegant Graphics for Data Analysis}.
\newblock Springer-Verlag New York.
\newblock ISBN 978-3-319-24277-4.
\newblock \urlprefix\url{https://ggplot2.tidyverse.org}.

\bibitem[{Wouk \emph{et~al.}(2019)Wouk, Bauer, and Gottfredson}]{Wouk2019}
Wouk K, Bauer AE, Gottfredson NC (2019).
\newblock \enquote{How to Implement Directed Acyclic Graphs to Reduce Bias in
  Addiction Research.}
\newblock \emph{Addictive Behaviors}, \textbf{94}, 109--116.
\newblock \doi{10.1016/j.addbeh.2018.09.032}.

\bibitem[{Zhang(2018)}]{Zhang2018}
Zhang X (2018).
\newblock \enquote{Application of Discrete Event Simulation in Health Care: A
  Systematic Review.}
\newblock \emph{BMC Health Services Research}, \textbf{18}(687).
\newblock \doi{10.1186/s12913-018-3456-4}.

\bibitem[{Zhang \emph{et~al.}(2018)Zhang, Reinikainen, Adeleke, Pieterse, and
  Groothuis-Oudshoorn}]{Zhang2018a}
Zhang Z, Reinikainen J, Adeleke KA, Pieterse ME, Groothuis-Oudshoorn CGM
  (2018).
\newblock \enquote{Time-Varying Covariates and Coefficients in Cox Regression
  Models.}
\newblock \emph{Annals of Translational Medicine}, \textbf{6}(7), 1--10.
\newblock \doi{10.21037/atm.2018.02.12}.

\end{thebibliography}


\newpage

\begin{appendix}

\section{Further Features of Discrete-Time Simulation} \label{appendix::dt_features}

As mentioned in Section~\ref{chap::dt_extentions}, there are a lot of possible features that could be included in the DGP using the discrete-time simulation approach, which are not shown in the main text. Below we will give some simple examples on how to use the discrete-time simulation approach implemented in the \pkg{simDAG} package to generate artificial data with some of these useful features. For most of these examples, the first DGP shown in Section~\ref{chap::covid_example} will be extended.

\subsection{Time-Dependent Base Probabilities} \label{sec::time_dep_prob}

To make $P_{Y0}$ time-dependent, the \code{sim\_time} argument may be used to change the definition of the function that generates the myocarditis probabilities. The following code gives an example for this:

\begin{Schunk}
\begin{Sinput}
R> prob_myoc <- function(data, P_0, RR_A, sim_time) {
+    P_0 <- P_0 + 0.001*sim_time
+    fifelse(data$A_event, P_0*RR_A, P_0)
+  }
R> dag <- empty_dag() +
+    node_td("A", type="time_to_event", prob_fun=0.01,
+            event_duration=20, immunity_duration=150) +
+    node_td("Y", type="time_to_event", prob_fun=prob_myoc,
+            parents=c("A_event"), P_0=0.005, RR_A=3.24)
R> sim <- sim_discrete_time(dag, n_sim=100, max_t=200)
R> data <- sim2data(sim, to="start_stop")
R> head(data)
\end{Sinput}
\begin{Soutput}
     .id start  stop      A      Y
   <int> <int> <num> <lgcl> <lgcl>
1:     1     1     7  FALSE  FALSE
2:     1     8     8  FALSE   TRUE
3:     1     9    31  FALSE  FALSE
4:     1    32    32  FALSE   TRUE
5:     1    33    57  FALSE  FALSE
6:     1    58    61   TRUE  FALSE
\end{Soutput}
\end{Schunk}

In this example, the base probability of myocarditis grows by 0.001 with each passing day, but the relative risk of infection given a vaccination stays constant. More complex time-dependencies may of course be implemented as well. The same procedure could also be used to make the vaccination probabilities time-dependent as well.

\subsection{Time-Dependent Effects} \label{sec::time_dep_effects}

In some scenarios it may also be required to make the effect itself vary over time. Here, we could make the size of the relative risk of developing a myocarditis given a recent Covid-19 vaccination dependent on the calender time, again by using the \code{sim\_time} argument in the respective \code{prob\_fun}:

\begin{Schunk}
\begin{Sinput}
R> prob_myoc <- function(data, P_0, RR_A, sim_time) {
+    RR_A <- RR_A + 0.01*sim_time
+    fifelse(data$A_event, P_0*RR_A, P_0)
+  }
R> dag <- empty_dag() +
+    node_td("A", type="time_to_event", prob_fun=0.01,
+            event_duration=20, immunity_duration=150) +
+    node_td("Y", type="time_to_event", prob_fun=prob_myoc,
+            parents=c("A_event"), P_0=0.005, RR_A=3.24)
R> sim <- sim_discrete_time(dag, n_sim=100, max_t=200)
R> data <- sim2data(sim, to="start_stop")
R> head(data)
\end{Sinput}
\begin{Soutput}
     .id start  stop      A      Y
   <int> <int> <num> <lgcl> <lgcl>
1:     1     1     5  FALSE  FALSE
2:     1     6    25   TRUE  FALSE
3:     1    26   193  FALSE  FALSE
4:     1   194   194  FALSE   TRUE
5:     1   195   200  FALSE  FALSE
6:     2     1    14  FALSE  FALSE
\end{Soutput}
\end{Schunk}

Here, the relative risk increases by 0.01 for each passing day, meaning that the risk of a myocarditis given a recent vaccination increases over time.

\subsection{Non-Linear Effects}

In all previous examples, it was assumed that the effect of the vaccination on the probability of developing a myocarditis follows a step-function. The risk was instantly elevated by a constant factor ($RR_A$) after vaccination, which lasts for a specified amount of time and then instantly drops back to the baseline risk. Any other kind of relationship may also be simulated, by again changing the \code{prob\_myoc()} function accordingly. The following code may be used:

\begin{Schunk}
\begin{Sinput}
R> prob_myoc <- function(data, P_0, RR_A) {
+    RR_A <- RR_A - 0.1*data$A_time_since_last
+    fifelse(data$A_event, P_0*RR_A, P_0)
+  }
R> dag <- empty_dag() +
+    node_td("A", type="time_to_event", prob_fun=0.01,
+            event_duration=20, immunity_duration=150,
+            time_since_last=TRUE) +
+    node_td("Y", type="time_to_event", prob_fun=prob_myoc,
+            parents=c("A_event", "A_time_since_last"),
+            P_0=0.005, RR_A=3.24)
R> sim <- sim_discrete_time(dag, n_sim=100, max_t=200)
R> data <- sim2data(sim, to="start_stop")
R> head(data)
\end{Sinput}
\begin{Soutput}
     .id start  stop      A      Y
   <int> <int> <num> <lgcl> <lgcl>
1:     1     1    10  FALSE  FALSE
2:     1    11    11  FALSE   TRUE
3:     1    12    64  FALSE  FALSE
4:     1    65    65  FALSE   TRUE
5:     1    66   200  FALSE  FALSE
6:     2     1   127  FALSE  FALSE
\end{Soutput}
\end{Schunk}

In this code, $RR_A$ decreases with each day after a person is vaccinated. In contrast to the previous example, this happens on a person-specific time-scale and not on a total calender time level. To do this properly, we set the \code{time\_since\_last} argument inside the \code{node\_td()} call of the vaccination definition to \code{TRUE}. This argument is supported when using nodes of type \code{"time\_to\_event"}. It adds another column to the dataset which includes the time since the last vaccination was performed for each individual. This column is then also added in the \code{parents} vector, so that we can use it in \code{prob\_myoc()} function. Here we simply substract 0.1 from $RR_A$ for each day after vaccination. This essentially means that on the day of vaccination itself, the relative risk will be 3.24 as specified in the \code{DAG}. On the first day after the vaccination, however, the relative risk will only be 3.14 and so forth. In other words, the extent of the adverse effect of the vaccination decreases over time linearly. Again, more complex functions may also be used to model this type of non-linear effects.

\subsection{Multiple Interrelated Binary Time-Dependent Variables}

Another possibility to extent the DGP would be to add more \code{"time\_to\_event"} variables. For example, we may want to additionally consider the effect of a Covid-19 infection itself, denoted by $C_t$ here. For simplicity we will assume that Covid-19 has a constant probability of occurrence over time which is the same for all individuals.In this example we will assume that the vaccination reduces the risk of getting a Covid-19 infection to 0 for $d_{immune}$ days after the vaccination was performed. We may use the following structural equation to describe this variable:

\begin{equation}
	C_t \sim Bernoulli(P_{C}(t)),
\end{equation}

with:

\begin{equation}
	P_{C}(t) =
	\begin{cases}
		0, & \text{if } t \in  \left[T_{A}(t), T_{A}(t) + d_{immune}\right] \\
		P_{C0}, & \text{otherwise}
	\end{cases},
\end{equation}

where $P_{C0}$ is the baseline probability of experiencing a Covid-19 infection and $T_A(t)$ is still defined to be the time of the last Covid-19 vaccination as before. In addition to this, we will also change the definition of the myocarditis node ($Y_t$). Instead of being only dependent on $A_t$, the Covid-19 Infection should now also raise the probability of developing a myocarditis by a constant factor $RR_C$ in the $d_{C.risk}$ days after the Covid-19 infection. The structural equation can then be changed to be:

\begin{equation}
	Y_t \sim Bernoulli(P_{Y}(t)),
\end{equation}

with:

\begin{equation}
	P_{Y}(t) =
	\begin{cases}
		P_{Y0} \cdot RR_{A} \cdot RR_{C}, & \text{if } t \in  \left[T_{A}(t), T_{A}(t) + d_{A.risk}\right] \text{ and } t \in  \left[T_{A}(t), T_{A}(t) + d_{C.risk}\right] \\
		P_{Y0} \cdot RR_{C}, & \text{if } t \in  \left[T_{A}(t), T_{A}(t) + d_{C.risk}\right] \\
		P_{Y0} \cdot RR_{A}, & \text{if } t \in  \left[T_{A}(t), T_{A}(t) + d_{A.risk}\right] \\
		P_{Y0}, & \text{otherwise}
	\end{cases},
\end{equation}

where $d_{A.risk}$ is the duration after vaccination in which the risk of developing a myocarditis is elevated by $RR_A$. The structural equation for $A$ are the same as defined in Equation~\ref{eq::vacc_bernoulli} and \ref{eq::vacc_pt}. The following code may be used to generate data from this DGP:

\begin{Schunk}
\begin{Sinput}
R> prob_myoc <- function(data, P_0, RR_A, RR_C) {
+    P_0 * RR_A^(data$A_event) * RR_C^(data$C_event)
+  }
R> prob_covid <- function(data, P_0, d_immune) {
+    fifelse(data$A_time_since_last < d_immune, 0, P_0, na=P_0)
+  }
R> dag <- empty_dag() +
+    node_td("A", type="time_to_event", prob_fun=0.01,
+            event_duration=20, immunity_duration=150,
+            time_since_last=TRUE) +
+    node_td("C", type="time_to_event", prob_fun=prob_covid,
+            parents=c("A_time_since_last"), event_duration=14,
+            P_0=0.01, d_immune=120) +
+    node_td("Y", type="time_to_event", prob_fun=prob_myoc,
+            parents=c("A_event", "A_time_since_last", "C_event"),
+            P_0=0.005, RR_A=3.24, RR_C=2.5)
R> sim <- sim_discrete_time(dag, n_sim=100, max_t=200)
R> data <- sim2data(sim, to="start_stop")
R> head(data)
\end{Sinput}
\begin{Soutput}
     .id start  stop      A      C      Y
   <int> <int> <num> <lgcl> <lgcl> <lgcl>
1:     1     1     8  FALSE  FALSE  FALSE
2:     1     9     9  FALSE  FALSE   TRUE
3:     1    10    18  FALSE  FALSE  FALSE
4:     1    19    38   TRUE  FALSE  FALSE
5:     1    39    39  FALSE  FALSE  FALSE
6:     1    40    40  FALSE  FALSE   TRUE
\end{Soutput}
\end{Schunk}

In this code we first re-define the \code{prob\_myoc()} function to include the effect of the Covid-19 infection. This utilizes a small computational trick, relying on the fact that any number to the power of 0 is 1 and any number to the power of 1 is itself. Since \code{TRUE} is treated as a 1 and \code{FALSE} is interpreted as a 0 in R, we can simply take the relative risks to the power of their current event indicator to only multiply the baseline risk with the respective relative risks if they are currently in the risk durations. Next, we define a new function called \code{prob\_covid()} to include the immunity duration, again using the \code{time\_since\_last} functionality. By finally adding another \code{node\_td()} call to the \code{DAG} object, we were able to specify the DGP in the desired way.

\subsection{Using Baseline Covariates}

Previously, only time-dependent variable were included in the DGP. By adding calls to the simple \code{node()} function to the \code{DAG} object it is, however, also possible to additionally include time-independent variables as well. Suppose that we want the baseline probability $P_{A0}$ of the vaccination to vary by biological sex in the first DGP described in Section~\ref{chap::covid_example}. We could do this using the following code:

\begin{Schunk}
\begin{Sinput}
R> prob_myoc <- function(data, P_0, RR_A) {
+    fifelse(data$A_event, P_0*RR_A, P_0)
+  }
R> prob_vacc <- function(data, P_0) {
+    fifelse(data$Sex, P_0*2, P_0)
+  }
R> dag <- empty_dag() +
+    node("Sex", type="rbernoulli", p=0.4) +
+    node_td("A", type="time_to_event", prob_fun=0.01,
+            event_duration=20, immunity_duration=150) +
+    node_td("Y", type="time_to_event", prob_fun=prob_myoc,
+            parents=c("A_event"), P_0=0.005, RR_A=3.24)
R> sim <- sim_discrete_time(dag, n_sim=100, max_t=200)
R> data <- sim2data(sim, to="start_stop")
R> head(data)
\end{Sinput}
\begin{Soutput}
     .id start  stop      A      Y    Sex
   <int> <int> <num> <lgcl> <lgcl> <lgcl>
1:     1     1    54  FALSE  FALSE   TRUE
2:     1    55    55  FALSE   TRUE   TRUE
3:     1    56   145  FALSE  FALSE   TRUE
4:     1   146   155   TRUE  FALSE   TRUE
5:     1   156   156   TRUE   TRUE   TRUE
6:     1   157   160   TRUE  FALSE   TRUE
\end{Soutput}
\end{Schunk}

Here, everything is kept the same as in the original example, with the small change that we now add a call to \code{node()} before adding the time-dependent variables to define the \code{Sex} node using a simple Bernoulli distribution. Additionally, we now had to define a function that appropriately generates the probabilities of vaccination per \code{Sex}. This was done by simply increasing $P_0$ by a factor of 2 whenever the value of \code{Sex} is \code{TRUE} (which might stand for males or females). It would also be possible to add child time-independent child nodes as well, but this is left as an exercise to the interested reader.

\subsection{Using Categorical Time-Dependent Variables}

The previous examples all included only binary time-dependent variables. For some applications it may be neccessary to used categorical time-dependent variables. For example, instead of generating the vaccination status as binary, we may want to model it as a categorical variable with multiple levels, indicating which kind of vaccine the person received (if any). This may be done using the \code{"competing\_events"} node \code{type}. Below is a simple example:

\begin{Schunk}
\begin{Sinput}
R> prob_myoc <- function(data, P_0, RR_A) {
+    fifelse(data$A_event > 0, P_0*RR_A, P_0)
+  }
R> prob_vacc <- function(data) {
+    n <- nrow(data)
+    p_mat <- matrix(c(rep(0.99, n), rep(0.0005, n),
+                      rep(0.0005, n)),
+                    byrow=FALSE, ncol=3)
+    return(p_mat)
+  }
R> dag <- empty_dag() +
+    node_td("A", type="competing_events", prob_fun=prob_vacc,
+            event_duration=c(20, 20), immunity_duration=150) +
+    node_td("Y", type="time_to_event", prob_fun=prob_myoc,
+            parents=c("A_event"), P_0=0.005, RR_A=3.24)
R> sim <- sim_discrete_time(dag, n_sim=100, max_t=200,
+                           save_states="all")
R> data <- sim2data(sim, to="start_stop")
R> head(data)
\end{Sinput}
\begin{Soutput}
     .id start  stop     A      Y
   <int> <int> <int> <num> <lgcl>
1:     1     1   200     0  FALSE
2:     2     1    85     0  FALSE
3:     2    86    86     0   TRUE
4:     2    87   200     0  FALSE
5:     3     1    95     0  FALSE
6:     3    96    96     0   TRUE
\end{Soutput}
\end{Schunk}

The \code{"competing\_events"} node \code{type} works in much the same way as the \code{"time\_to\_event"} node type, with the main difference being that instead of using Bernoulli trials it relies on multinomial trials, which are equivalent to drawing a simple random sample with unequal probabilities for each element. Instead of supplying a single probability of success per person and per point in time, the \code{prob\_fun} supplied to a \code{"competing\_events"} node is therefore expected to provide a vector of probabilities. In the example above, we simply define the probabilities to be the same for everyone regardless of the simulated time. The first category represents ``no vaccination'' while the next two categories specify vaccinations with vaccines of different kinds. Note that in this code, the \code{prob\_myoc()} function only checks whether there was any vaccination, so it does not really make a difference whether one uses a \code{"competing\_events"} or \code{"time\_to\_event"} node here.
\par\medskip
To get more useful person specific multinomial probabilities, the \code{node\_multinomial()} function may be useful. Because the underlying code for multinomial trials is a bit more complex than the code for simple Bernoulli trials, using this type of node may lead to an increased runtime. The node type is usually most useful when the goal is to generate artificial time-to-event data with mutually exclusive types of events, also known as competing events or competing risks data. Also note that in the example we had to set the \code{save\_states} argument of the \code{sim\_discrete\_time()} function to \code{"all"}. This is required when the goal is to transform the data into the start-stop format later, because the DGP does not consist solely of time-dependent nodes of type \code{"time\_to\_event"}.

\subsection{Using Continuous Time-Dependent Variables}

It is also possible to use continuous variables as time-dependent variables in the proposed package. The following code gives a very simple example:

\begin{Schunk}
\begin{Sinput}
R> dag <- empty_dag() +
+    node("calories", type="rnorm", mean=2500, sd=150) +
+    node_td("calories", type="gaussian",
+            formula= ~ 1 + calories*1.1, error=1)
R> sim <- sim_discrete_time(dag, n_sim=100, max_t=200,
+                           save_states="all")
R> data <- sim2data(sim, to="long")
R> head(data)
\end{Sinput}
\begin{Soutput}
Key: <.id, .time>
     .id .time calories
   <int> <int>    <num>
1:     1     1 2687.018
2:     1     2 2958.310
3:     1     3 3254.088
4:     1     4 3581.853
5:     1     5 3942.032
6:     1     6 4335.582
\end{Soutput}
\end{Schunk}

In this example, we first generate a normally distributed root node called \code{calories} using a standard \code{node()} call. This represents the value of the variable at $t = 0$. If we did not specify this variable, the code would return an error message at $t = 1$, because there would be no value of \code{calories} to use in the subsequently defined regression model. Next, a call to \code{node\_td()} is added to the \code{DAG} object to specify how this variable changes with each step in time. We use a simple linear regression model, where the only independent variable is the last value of the variable itself. In the subsequent simulation we again set \code{save\_states="all"} to specify that the data at all points in time should be saved. This is necessary here both because the \code{DAG} consists of time-dependent nodes that are not of type \code{"time\_to\_event"}, and because the variable changes at every point in time for every individual. Because of the nature of the data, it would also not make sense to transform the data into the start-stop format. Instead, the long format is choosen here.

\subsection{Ordered Events}

The following code gives an example on how events could be simulated so that a specific order of events is always respected:

\begin{Schunk}
\begin{Sinput}
R> prob_bachelors <- function(data) {
+    fifelse(data$highschool_event, 0.01, 0)
+  }
R> prob_masters <- function(data) {
+    fifelse(data$bachelors_event, 0.01, 0)
+  }
R> dag <- empty_dag() +
+    node_td("highschool", type="time_to_event", prob_fun=0.01,
+            event_duration=Inf) +
+    node_td("bachelors", type="time_to_event", prob_fun=prob_bachelors,
+            event_duration=Inf, parents="highschool_event") +
+    node_td("masters", type="time_to_event", prob_fun=prob_masters,
+            event_duration=Inf, parents="bachelors_event")
R> sim <- sim_discrete_time(dag, n_sim=100, max_t=200)
R> data <- sim2data(sim, to="start_stop")
R> head(data)
\end{Sinput}
\begin{Soutput}
     .id start  stop highschool bachelors masters
   <int> <int> <num>     <lgcl>    <lgcl>  <lgcl>
1:     1     1    19      FALSE     FALSE   FALSE
2:     1    20    82       TRUE     FALSE   FALSE
3:     1    83    99       TRUE      TRUE   FALSE
4:     1   100   200       TRUE      TRUE    TRUE
5:     2     1    75      FALSE     FALSE   FALSE
6:     2    76   200       TRUE     FALSE   FALSE
\end{Soutput}
\end{Schunk}

In this specification, we try to simulate the educational status of a person over time. We assume that individuals can only obtain a bachelors degree once they have finished high school and that they can only receive a masters degree once they finished the bachelors degree. To keep this order of events in tact, we can split the variable into its three constituent parts. First, the indicator whether someone graduated \code{highschool} is simulated using simple Bernoulli trials as implemented in the \code{"time_to_event"} node type. By setting \code{event_duration=Inf}, we ensure that no one ever "looses" their degree. Only afterwards do we generate whether the same person also received a \code{bachelors} degree. By using a standard \code{fifelse()} call, we can easily specify that the probability of obtaining a \code{bachelors} degree is 0 for individuals without a \code{highschool} degree. Subsequently, we do the same for the \code{masters} node.
\par
Again, this simulation could be made much more realistic. For example, in the example used here there is no limitations on how close the different graduations can be to each other. An individual might receive all three degrees in the same time unit. This could be changed by using the \code{sim_time} argument in the definition of the probability functions, as discussed in Section~\ref{sec::time_dep_prob} and Section~\label{sec::time_dep_effects}.

\end{appendix}


\end{document}